\documentclass[preprint,unsortedaddress]{revtex4-1}
\usepackage{graphicx}
\usepackage{epstopdf}
\usepackage{subfig,float}
\usepackage{amsmath,amssymb}
\usepackage{natbib}
\usepackage{xcolor}
\usepackage{dcolumn}%
\usepackage{bm}%
\newcommand{\lra}[1]{\langle #1 \rangle }

\newcommand{\mc}[1]{\mathcal{#1}}

\newcommand{\mb}[1]{\mathbf{#1}}
\newcommand{\bds}[1]{\boldsymbol{#1}}

\def\be{\begin{equation}}
\def\ee{\end{equation}}

\draft 

\begin{document}


\title{Lagrangian filtered density function for LES-based stochastic modelling of turbulent dispersed flows} 



\author{Alessio Innocenti}
\affiliation{Sorbonne University, UPMC Univ Paris 06, CNRS, UMR 7190, Institut Jean Le Rond d'Alembert, F-75005 Paris, France}
\affiliation{Dipartimento di Ingegneria Civile e Industriale, Universit\`a di Pisa, Via G. Caruso 8, 56122 Pisa, Italia}

\author{Cristian Marchioli}
\affiliation{Department of Fluid Mechanics, University of Udine, 33100 Udine, Italy}

\author{Sergio Chibbaro}
\affiliation{Sorbonne University, UPMC Univ Paris 06, CNRS, UMR 7190, Institut Jean Le Rond d'Alembert, F-75005 Paris, France}



\begin{abstract}
The Eulerian-Lagrangian approach based on Large-Eddy Simulation (LES) is one of the most
promising and viable numerical tools to study turbulent dispersed flows when the computational cost of
Direct Numerical Simulation (DNS) becomes too expensive. The applicability of this approach is however
limited if the effects of the Sub-Grid Scales (SGS) of the flow on particle dynamics are neglected.
In this paper, we propose to take these effects into account
by means of a Lagrangian stochastic SGS model for the equations of particle motion. The model extends to
particle-laden flows the velocity-filtered density function method originally developed for reactive flows.
The underlying filtered density function is simulated through a Lagrangian Monte Carlo procedure
that solves for a set of Stochastic Differential Equations (SDEs) along individual particle trajectories.
The resulting model is tested for the reference case of turbulent channel flow,
using a hybrid algorithm in which the fluid velocity field is provided by LES and then used to advance
the SDEs in time. The model consistency is assessed in the limit of particles
with zero inertia, when ``duplicate fields'' are available from both the
Eulerian LES and the Lagrangian tracking. Tests with inertial particles were performed to examine
the capability of the model to capture particle preferential concentration and near-wall segregation.
Upon comparison with DNS-based statistics, our results show improved accuracy and considerably
reduced errors with respect to the case in which no SGS model is used in the equations of particle motion.
\end{abstract}


\maketitle 


\section{Introduction}

Over the past decades major modelling efforts have been devoted to the prediction of single-phase turbulent flows by means of Large Eddy Simulation (LES) \cite{rogallo1984numerical,sagaut2006large,lesieur2005large}.	
The pioneering model was developed by Smagorinsky \cite{smagorinsky1963general}, based on an eddy viscosity closure that relates the unknown Sub-Grid Scale (SGS) stresses to the strain rate of the large flow scales to mimic the dissipative behavior of the unresolved flow scales. Subsequent extensions to dynamic \cite{germano1991dynamic,germano1992turbulence} or stochastic models \cite{zamansky2013acceleration}
have improved the quality and reliability of LES, especially for cases where mass, heat and momentum transfer
are controlled by the large scales of the flow. 
Much work has been done also to improve the applicability of LES to chemically-reacting turbulent flows
\cite{Fox2003,pope2013small} and, more recently, to dispersed turbulent flows \cite{fox2012large}.
The first LES of particle-laden flow, in particular, was performed under the assumption of negligible contribution
of the SGS fluctuations to the filtered fluid velocity seen by inertial particles \cite{armenio1999effect}: The choice
was justified considering that inertial particles act as low-pass filters that respond selectively to removal of SGS flow
scales according to a characteristic frequency proportional to $1/\tau_p$, where $\tau_p$ is the particle relaxation time
(a measure of particle inertia). The same assumption has been used in other studies
\cite{Yam_01,Van_06,Dri_11,Afk_15} in which the filtering due to particle inertia
and the moderate Reynolds number of the flow had a relatively weak effect on the (one-particle, two-particles)
dispersion statistics examined.
However, several studies \citep{Kue_06,Mar_08,Calzavarini2010} have demonstrated that neglecting the effect of
SGS velocity fluctuations on particle motion leads to significant errors in the quantification of large-scale clustering
and preferential concentration, two macroscopic phenomena that result from particle preferential distribution
at the periphery of strong vortical regions into low-strain regions \citep{fessler1994preferential,Wan_93,Rou_01}.
It is now well known that LES without SGS modelling for the dispersed phase is bound to underestimate
preferential concentration and, in turn, deposition fluxes and near-wall accumulation \cite{Sol_12AWR,Sol_09,Pic_05}. 
These flaws have
obvious consequences on the applicability of LES to industrial processes and environmental phenomena such
as mixing, combustion, depulveration, spray dynamics, pollutant dispersion, or cloud dynamics \cite{Bala_10}.
Recent analyses based on Direct Numerical Simulation (DNS) of turbulence have also shown that neither
deterministic models nor stochastic homogeneous models have the capability to correct fully the inaccuracy
of the LES approach due to SGS filtering \cite{Mar_08ACME,bianco2012intrinsic,geurts2012ideal,chibbaro2014particle}.
Prompted by the above-mentioned findings, some attempts have been made on a heuristic ground, 
both for isotropic  \citep{Poz_09,Gob_10,Sho_05,cernick2015particle} and wall-bounded flows
\cite{michalek2013hybrid,jin2015simple}.

An interesting and viable modelling alternative is represented by the Probability Density Function (PDF) approach, which
has proven useful for LES of turbulent reactive flows \cite{Col_98,Jab_99,gicquel2002velocity,She_03,She_07,She_09}.
The LES formalism is based on the concept of Filtered Density Function (FDF), which is essentially the filtered
fine-grained PDF of the transport quantities that characterize the flow. 
In this framework, the SGS effect is
included in a set of suitably-defined Stochastic Differential Equations (SDEs), where the effects of advection,
drag non-linearity and poly-dispersity appear in a closed form. This constitutes the primary advantage of the PDF/FDF approach with respect to other statistical procedures,
in which these effects require additional modelling \cite{pope2000turbulent}. 

The objective of the present work is to develop the FDF-based LES formalism for particle-laden
turbulent flows. To this aim, several issues must be addressed with respect to the FDF approach already
available for turbulent reactive flows. First, the FDF must be Lagrangian since particle dynamics are addressed naturally from the Lagrangian viewpoint. In addition, inertial particles behave like a compressible
phase  and
therefore the mass density function should be considered. This leads to the definition of
a joint Lagrangian Filtered Mass Density Function (LFMDF), which represents the mathematical framework
required to implement the FDF approach in LES.
In particular, a suitable transport equation must be developed for the LFMDF such that the effects of SGS
convection appear in closed form (the unclosed terms in the transport equation can be modelled following
a procedure similar to Reynolds averaging). In this paper, the numerical solution of the LFMDF transport equation
is achieved by means of a Lagrangian Monte Carlo procedure. The consistency of this procedure is assessed by comparing the first two moments of the LFMDF with those obtained from the Eulerian LES of the flow.
The results provided by the LFMDF simulations are compared with those predicted by the original
Smagorinsky closure, as well as those of the ``dynamic'' Smagorinsky model, for the reference case of turbulent
channel flow.
The LFMDF performance is further assessed upon direct comparison with a DNS dataset, paying particular attention
to the results for particle preferential concentration.

\vfill

\section{Problem Formulation}

In the mathematical description of turbulent dispersed flows, the relevant transport variables are the fluid
velocity $U_i({\bf x},t)$,
the pressure $P$, the particle position ${\mb x_p(t)}$, and the particle velocity ${\mb U}_p({\bf x}_p(t),t)$.
In this work, we consider heavy particles carried by an incompressible Newtonian fluid.
The equations of motion for the fluid are, in scalar form:
\begin{eqnarray}
\label{fluid: exact Cont}
 \frac{\partial U_i}{\partial x_i}&=&0 ~,\\
 \label{fluid: exact NS}
\frac{\partial U_i}{\partial t} &+& U_j\frac{\partial U_i}{\partial x_j} = 
  -\frac{1}{\rho_f}\frac{\partial P}{\partial x_i} + 
   \nu_f \frac{\partial^2 U_i}{\partial x_j^2} ~,
\end{eqnarray}
where $\rho_f$ and $\nu_f$ are the density and the kinematic viscosity of the fluid, respectively.
LES of turbulence involves the use of a spatial filter \cite{germano1992turbulence}:
\begin{equation}
\label{filtering-operator}
\widetilde{f}({\mb x},t)=\int_{-\infty}^{\infty} f({\bf y},t)G({\bf y},{\bf x})d{\bf y}~,
\end{equation}
where $G$ is the filter function, $\widetilde{f}$ represents the
filtered value of the transport variable $f$, and $f^{\prime}=f-\widetilde{f}$
denotes the fluctuation of $f$ with respect to the filtered value. We consider spatially- and
temporally-invariant, localized filter functions, thus $G (y,x)\equiv G(x-y)$ with the
properties, $G(x)=G(-x)$, and $\int G(x)dx=1$.
Starting from Eqns. (\ref{fluid: exact Cont}) and (\ref{fluid: exact NS}) , application of the filtering
operator (\ref{filtering-operator}) yields:
 \begin{eqnarray}
 \label{eq:les}
 \frac{\partial \widetilde{U}_j}{\partial x_j}&=&0~, \\
\frac{\partial \widetilde{U_i}}{\partial t}&+&\widetilde{U}_j\frac{\partial \widetilde{U_i}}{\partial x_j}=
-\frac{1}{\rho_f}\frac{\partial \widetilde{P}}{\partial x_i} +\nu_f  \frac{\partial^2 \widetilde{ U_i} }{\partial x_j^2} 
-\frac{\partial \widetilde{\tau}_{ij}}{\partial x_j}~,
\end{eqnarray}
where $\widetilde{\tau}_{ij}=\widetilde{U_i U_j}- \widetilde{U}_i \widetilde{U}_j$
is the SGS tensor component \cite{germano1992turbulence}.
To close the SGS stress tensor, three different cases have been considered in order to compare the differences
produced on particle tracking: $(1)$ no SGS model, $(2)$ Smagorinsky SGS model \cite{smagorinsky1963general} and $(3)$  Germano (dynamic Smagorinsky) SGS model 
\cite{germano1991dynamic,germano1992turbulence,lilly1992proposed}.
In the case without SGS model, the contribution of the SGS is completely ignored
and $\widetilde{\tau}_{i,j} = 0$. The Smagorinsky model reads \cite{smagorinsky1963general}:
\begin{equation}
\widetilde{\tau}_{i,j} - \frac{2}{3} k \delta_{i,j} = - 2 \nu_t \widetilde{S}_{i,j}~,
\end{equation}
\begin{equation}
\widetilde{S}_{i,j} = \frac{1}{2} \Bigl ( \frac{\partial \widetilde{U_i}}{\partial x_j} + \frac{\partial \widetilde{U}_j}{\partial x_i} \Bigl )~,
\end{equation}
\begin{equation}
\nu_t = (C_S \Delta)^2 \mathcal{S}~,
\end{equation}
with $C_S = 0.065$ \cite{moin1982numerical}, $ \mathcal{S} =\sqrt{\widetilde{S}_{i,j} \widetilde{S}_{i,j}}$ and $\Delta$ the characteristic length of the filter. The dynamic version of the Smagorinsky model provides a means of approximating $C_S$ (the reader is referred to \cite{germano1992turbulence} for further details on the model).

For the case of heavy particles (with density $\rho_p \gg \rho_f$), drag 
and gravity are the dominant forces and the equations of particle motion in
the Lagrangian framework, and in vector form, read as \cite{Cli_78}:
\begin{align}  \label{pospar}
\frac{d \mb{x}_{p}}{dt} &= {\mb{U}_{p}}~, \\
  \label{velpar}
  \frac{d\mb{U}_{p}}{dt} &= \frac{1}{\tau_{p}}(\mb{U}_{s}-\mb{U}_{p}) + \mb{g}~,
\end{align}
where $\mathbf{U}_s=\mathbf{U}(\mathbf{x}_p,t)$ is the fluid velocity seen by
a particle along its trajectory, and:
\begin{equation} \label{definition taup}
\tau_{p}=\frac{\rho_p}{\rho_f}\frac{4\,d_p}{3\,C_D |\mb{U}_r |}~,
\end{equation}
is the particle relaxation time, with $d_p$ the particle diameter, $C_D=\frac{24}{Re_p}(1 + 0.15 Re_p^{0.687})$
the drag coefficient and $\mb{U}_r=\mb{U}_p-\mb{U}_s$ the particle-to-fluid relative velocity at the
particle position.
Similarly to what already done for the fluid phase, it is possible to derive the filtered version
of Eqns. (\ref{pospar}) and (\ref{velpar}). The Lagrangian nature of these equations, however,
does not allow a straightforward derivation unless the SGS effects on particle motion are
disregarded. In this case one can write:
\begin{align}
\frac{d\widetilde{\mb{x}}_{p}}{dt} &= \widetilde{\mb{U}}_{p}~, \\
\frac{ d\widetilde{\mb{U}}_{p}}{dt} &= \frac{\widetilde{\mb{U}}_s - \widetilde{\mb{U}}_p}{\widetilde{\tau}_p} + \mb{g}~,
\end{align}
where $\widetilde{\tau}_p$ is the particle relaxation timescale expressed in 
terms of the filtered relative velocity $\widetilde{\mb{U}}_r$. 
A more precise definition of the filtering procedure for the particle-phase quantities
is given in the following section.

\section{ Definition of the Filtered Density Function}
\label{PDF methods}

\subsection{Particle phase}

In polydispersed two-phase flows the exact governing equations are Lagrangian. Accordingly,
we introduce a  \textit{Lagrangian Filtered Mass Density Function}  (LFMDF) that is formally
defined for $N$ individual particles in the domain at the time $t$ as: 
\begin{eqnarray}
\label{eq:Ftild2}
\widetilde{F}^{p}_L(t;\mb{y}_p,\bds{V}_p,\bds{V}_s)&= &
\int \sum_{i=1}^N  m_{p,i}\, G({\bf y}-{\bf y'}_p)\delta({\bf y'}_p-{\bf x}_{p,i}(t))\otimes
\delta(\bds{V}_p-{\bf U}_{p,i}(t)) \otimes \delta({\bds V}_s - {\bf U}_{s,i}(t))d{\bf y'}\notag \\
&=&\sum_{i=1}^N m_{p,i}\, G({\bf y}-{\bf x}_{p,i}(t))\otimes 
\delta({\bds V}_p - {\bf U}_{p,i}(t)) \otimes \delta({\bds V}_s - {\bf U}_{s,i}(t))~,
\end{eqnarray}
where $m_{p,i}$ is the mass of the $i$-th particle.
From the LFMDF, it is possible to derive formally the corresponding Eulerian Filtered Mass Density Function
(EFMDF):
\begin{eqnarray}
\nonumber
 {\widetilde F}_E^p(t,\mb{x};\bds{V}_p,\bds{V}_s) \equiv\, \widetilde{F}_L^p(t;\mb{y}_p=\mb{x},\bds{V}_p,\bds{V}_s)  =\\
\label{eq:FE1}
= \sum_{i=1}^N m_{p,i}\, G({\bf x}-{\bf x}_{p,i}(t))\otimes \delta({\bds V}_p - {\bf U}_{p,i}(t)) \otimes \delta({\bds V}_s - {\bf U}_{s,i}(t))~.
\end{eqnarray}
Let us now consider the conditional filtered value of a variable $Q(t)$, which is defined as follows:
\begin{equation}
\label{eq:condfilt}
\lra{\widetilde{Q}(t) |{\bf  y}_p,\bds{V}_p,\bds{V}_s} 
=\frac{\sum_{i=1}^N Q_i   m_{p,i} G({\bf y}_p-{\bf x}_{p,i})\otimes 
\delta({\bds V}_p-{\bf U}_{p,i}(t)) \otimes \delta(\bds{V}_s-{\bf U}_{s,i}(t))}{\widetilde{F}^{p}_L(t;\mb{y},\bds{V}_p,\bds{V}_s)}~.
\end{equation}
Equations (\ref{eq:FE1}) and (\ref{eq:condfilt}) imply that:
\begin{enumerate}
\item[(i)] if $Q(t)=const.$ then $\lra{\widetilde{Q}(t) |{\bf  y},\bds{V}_p,\bds{V}_s}=const.$
\item[(ii)] if $Q(t)\equiv \hat{Q}(\bds{x}(t), \bds{U}_p(t), \bds{U}_{s}(t))$, namely when the variable $Q$ is completely
defined by the variables $ \bds{x}(t)$, $\bds{U}_p(t)$, and $\bds{U}_{s}(t)$, then
$\lra{\widetilde{Q}(t) |{\bf  y},\bds{V}_p,\bds{V}_s}=\hat{Q}({\bf y},\bds{V}_p,\bds{V}_s)$
 \item[(iii)] the following integral property for any variable $Q(t,\mb{x})$ holds:
\begin{equation} \label{Eulerian Filtered field}
\alpha_{p}(t,\mb{x})\lra{\rho}_p \widetilde{Q} (t,\mb{x}) =
\int \int \lra{\widetilde{Q} |{\bf  y}={\bf  x},\bds{V}_p,\bds{V}_s}
\;\; \widetilde{F}_E^p(t,\mb{x};\bds{V}_p,\bds{V}_s)
\,d\mb{V}\,d\bds{U_s}~,
\end{equation}
\end{enumerate}
where $\alpha_{p}(t,\mb{x})\lra{\rho}_p=
\int \widetilde{F}_E^p(t,\mb{x};\mb{V}_p,\bds{V}_{s})
\,d\mb{V}_p\,d\bds{V}_{s}$ is the filtered local value of the particle mass 
fraction at time $t$ and position $\mb{x}$.
From these equations, it follows that the filtered value of any function of the variables
in the state-vector is obtained by integration in the sample space:
\begin{equation} 
\alpha_{p}(t,\mb{x})\lra{\rho}_p \widetilde{Q} (t,\mb{x}) =
\int \int \hat{Q}( \bds{V}_p, \bds{V}_s)
\;\; \widetilde{F}_E^p(t,\mb{x};\mb{V},\bds{U}_s)
\,d\mb{V}\,d\bds{U}_s~.
\end{equation}

\subsection{LFMDF transport equation}
To derive the LFMDF transport equation, we consider the time derivative
of the  fine-grained density function given by Eq. (\ref{eq:Ftild2}).
Assuming that all particles have the same mass (namely $m_{p,i}$ is the same for $i=1, ... , N$
as for mono-dispersed flows), we can derive:
\begin{align}
\frac{\partial \widetilde{F}^{p}_L}{\partial t} &= 
\sum_{i=1}^N \left(m_{p,i} \frac{\partial G}{\partial t}\delta_{V_p V_s}+m_{p,i} G \frac{\partial \delta_{V_p}}{\partial t}\delta_{V_s}+
m_{p,i} G \frac{\partial \delta_{V_s}}{\partial t}\delta_{V_p}\right) \notag \\
&=\sum_{i=1}^N \left(m_{p,i} \frac{ \partial G}{\partial \mathbf{x}}\frac{d \mathbf{x}_i}{d t}
\delta_{V_p U_s}
- m_{p,i} G \frac{d \mathbf{U}_{p,i}}{d t}\frac{\partial \delta_{V_p}}{\partial {\mathbf V}_p}\delta_{U_s}-
m_{p,i} G \frac{d {\mathbf U}_{s,i}}{d t} \frac{\partial \delta_{V_s}}{\partial {\bf V}_s} \delta_{{V_p}} \right) \notag \\
&=\sum_{i=1}^N 
\left(-m_{p,i} \frac{ \partial G}{\partial \mathbf{y}}\frac{d \mathbf{x}_i}{d t}\delta_{V_p U_s}-
m_{p,i} G \frac{d \mathbf{U}_{p,i}}{d t}\frac{\partial \delta_{V_p}}{\partial {\mathbf V}_p}\delta_{U_s}-
m_{p,i} G \frac{d \mathbf{U}_{s,i}}{d t} \frac{\partial \delta_{V_s}}{\partial {\bf V}_s} \delta_{{V_p}} \right) \notag \\
&=\sum_{i=1}^N 
\left( -\frac{ \partial }{\partial \mathbf{y}}(m_{p,i}G \frac{d \mathbf{x}_i}{d t}\delta_{V_p U_s})-
\frac{\partial}{\partial \mathbf{V_p}}(m_{p,i} G \frac{d \mathbf{U}_{p,i}}{d t} \delta_{V_p} \delta_{U_s})-
\frac{\partial }{\partial {\bf V}_s}(m_{p,i} G \frac{d \mathbf{U}_{s,i}}{d t}  \delta_{{V_p}}\delta_{V_s}) \right) \notag \\
&=-\frac{ \partial }{\partial \mathbf{y}}
\left [\, \left \langle \widetilde{\frac{d \mathbf{x}}{d t}} |{\bf  y},\bds{V}_p,\bds{U}_s\right \rangle \,
\widetilde{F}_L^p \, \right ] 
-\frac{\partial}{\partial \mathbf{V}_p}
\left [ \left \langle \widetilde{\frac{d {\mathbf U}_p}{d t}} |{\bf  y},\bds{V}_p,\bds{U}_s\right \rangle \, \widetilde{F}_L^p \right ]
-\frac{\partial }{\partial {\bf V}_s}
\left [\,  \left \langle \widetilde{ \frac{d {\mathbf U}_s}{d t}} | {\bf  y},\bds{V}_p,\bds{V}_s \right \rangle \,
\widetilde{F}_L^p \, \right ] \notag \\
&=-\frac{\partial [\, \mathbf{V}_p \widetilde{F}_L^p \,]}{\partial \mathbf{y}}
-\frac{\partial }{\partial {\mathbf V}_p} 
\left [-\frac{{\mb V}_p - {\mb V}_s}{\tau_p} \, \widetilde{F}_L^p \right ]
-\frac{\partial }{\partial {\bf V}_s}
\left [\, \left \langle \widetilde{ \mathbf{A}}_{U_s} | {\bf  y},\bds{V}_p,\bds{U}_s\right \rangle \,
\widetilde{F}_L^p \, \right ]~.
\label{eq:LFMDF}
\end{align}
The LFMDF transport equation can be also written separating the filtered and unresolved parts as follows:
\begin{align}
\frac{\partial \widetilde{F}^{p}_L}{\partial t} +
\frac{\partial \left (\, \widetilde{\mb{U}}_p \widetilde{F}_L^p \, \right )}{\partial \mathbf{y}} =&
-\frac{\partial }{\partial \mb{V}_p} \left [\, \widetilde{\mb{A}}_{U_p}\, \widetilde{F}_L^p \, \right ]
-\frac{\partial }{\partial \mb{V}_s} \left [\, \widetilde{\mathbf{A}}_{U_s} \widetilde{F}_L^p \, \right ] \notag \\
& - \frac{\partial}{\partial \mathbf{y}}
\left [\, \left(\mathbf{V}_p - \widetilde{\mb{U}}_p\right) \widetilde{F}_L^p \, \right ] \notag \\
& -\frac{\partial }{\partial \mathbf{V}_p} 
\left \{
\left [
\left \langle
\widetilde{ \mathbf{A}}_{U_p} | {\bf  y},\bds{V}_p,\bds{V}_s
\right \rangle
- \widetilde{\mathbf{A}}_{U_p}
\right ]
\widetilde{F}_L^p
\right \}  \notag \\
&-\frac{\partial }{\partial \mb{V}_s}
\left \{
\left [
\left \langle
\widetilde{ \mathbf{A}}_{U_s} | {\bf  y},\bds{V}_p,\bds{V}_s
\right \rangle
- \widetilde{\mathbf{A}}_{U_s}
\right ]
\widetilde{F}_L^p
\right \}~,
\end{align}
where the first term on the right-end side corresponds to the effects of resolved scales whereas 
the last three terms take into account the effects of the unresolved scales.
The EFMDF $\widetilde{F}_E^p$ follows by definition the same transport equation.
%
%
%
%
\subsection{Modeled LFMDF transport equation}
The Langevin model developed previously for poly-dispersed flows in the RANS context~\cite{Min_01,Min_04,minier2015lagrangian}
is employed here to close the LFMDF transport equation. The modeled LFMDF equation reads as:
\begin{eqnarray}
&-&\frac{\partial }{\partial {\bf V}_{s}}
\left [\, \left \langle \widetilde{ \mathbf{A}}_{U_s} | {\bf  y},\bds{V_p},\bds{U_s}\right \rangle \,
\widetilde{F}_L^p \,\right ] \notag \\ 
&\approx&
-  \frac{\partial }{\partial {V}_{s,i}}
\left \{ \left [
-\frac{1}{\rho_f}\frac{\partial \widetilde{P}}{\partial x_i}
+\nu_f \Delta \tilde{U}_i
+ \left( \widetilde{U}_{p,j} - \widetilde{U}_{j} \right) 
\frac{\partial \widetilde{U}_{i}}{\partial x_j}
- \frac{V_{s,i} - \widetilde{U_{i}}}{T_{L,i}^{*}} \right ]\,
  \widetilde{F}_L^p \, \right \} \notag \\
&&+\frac{1}{2}\frac{\partial^2}{\partial V_{s,i}^2} 
\left \{\, \tilde{\epsilon}  \left [ C_0 b_i \frac{\widehat{k}_{SGS}}{k_{SGS}} + \frac{2}{3} \Bigl ( b_i \frac{\widehat{k}_{SGS}}{k_{SGS}} - 1 \Bigl ) \right ] 
\, \widetilde{F}_L^p \, \right \}~,
\label{eq:fp}
\end{eqnarray}
where we have defined the Lagrangian timescale in
the longitudinal direction ($i=1$), and in the transversal directions ($i=2$ and $i=3$
respectively) as:\\
\begin{equation}
\label{eq:CsanadyL}
T_{L,1}^{*}= \frac{T_{SGS}} {\sqrt{ 1 + \beta^2 \displaystyle \frac{\vert
      \widetilde{{\bf U}_r}\vert^2}{2k_{SGS}/3}}}~,~~~T_{L,2}^{*}= T_{L,3}^{*} = \frac{T_{SGS}} {\sqrt{ 1 + 4\beta^2 \displaystyle
    \frac{\vert \widetilde{{\bf U}_r}\vert^2}{2k_{SGS}/3}}}~,
\end{equation}
with $\beta = T_L / T_E$ \cite{wang1993dispersion}, and:
\begin{equation}
\tilde{\epsilon}=(C_S \Delta)^2 \mathcal{S} ~,\;\; k_{SGS}=C_{\epsilon} (\Delta \tilde{\epsilon})^{2/3} ~,\;\; T_{SGS}=\frac{k_{SGS}}{\tilde{\epsilon}}
\left ( \frac{1}{2}+\frac{3}{4}C_0  \right)^{-1}~,
\label{eq:def-eps}
\end{equation}
where $\tilde{\epsilon}$ is the SGS dissipation rate, $\Delta$ is the filter width, $k_{SGS}$ is the SGS kinetic energy, and $T_{SGS}$ is the SGS time-scale.
This model is consistent with the Generalised Langevin Model \cite{pope2000turbulent}.
The auxiliary subgrid turbulent kinetic energy is defined as follows:  
\begin{equation}
\widehat{k}_{SGS}= \frac{3}{2} \frac{\sum^3_{i=1}b_i [\widetilde{U_{s,i}^2} - \widetilde{ U_{s,i} \widetilde{U}_{s,i} } ]}{\sum^3_{i=1}b_i}~,
\label{eq:kappa}
\end{equation}
with $b_i=T_{SGS} / T_{L,i}^{*}$.

\subsection{Equivalent Stochastic System}
\label{sec:equiv_stoc_syst}
The LFMDF transport equation is of the Fokker-Planck kind and provides all the statistical information of the state-vector. 
However, the most convenient way to solve this equation is through a Lagrangian Monte
Carlo method, since the LFMDF equation is equivalent to a system of SDEs
in a weak sense \cite{Gar_90}.
This approach applies naturally to the dispersed phase since its evolution equations are Lagrangian.
The system of SDEs corresponding to Eq. (\ref{eq:fp}) reads as:
\begin{align}
& dx_{p,i}= U_{p,i}\, dt~, \label{eq:sdeXp}\\
& dU_{p,i} = \frac{U_{s,i} - U_{p,i}}{\tau_p}\, dt~, \label{eq:sdeUp}\\
& dU_{s,i} = -\frac{1}{\rho_f}\frac{\partial \widetilde{P}}{\partial x_i}\, dt
+\nu_f \Delta \tilde{U}_i
+\left( \widetilde{U}_{p,j} - \widetilde{U}_{j} \right) 
\frac{\partial \widetilde{U_i}}{\partial x_j}\, dt
- \frac{U_{s,i} - \widetilde{U}_{i}}{T_{L,i}^{*}}\, dt 
+ B_{s,ij} \, dW_i~,
\label{eq:sde}
\end{align}
where the term $dW_i$ denotes a Wiener process \cite{Gar_90}.
In the following we discuss the results obtained with two choices for the
 diffusion matrix $B_{s,ij}=\sqrt{C_i^{*} \widetilde{\epsilon}}\;\delta_{ij}$: 
 \begin{enumerate}
\item the complete model $C_i^{*}=\left [ C_0 b_i \frac{\widehat{k}_{SGS}}{k_{SGS}} + \frac{2}{3} \Bigl ( b_i \frac{\widehat{k}_{SGS}}{k_{SGS}} - 1 \Bigl ) \right ]$,  referred to as \textit{LFMDF2} hereinafter;
\item a simplified model $C_i^{*}\approx C_0\, b_i + \frac{2}{3}(b_i -1) $, referred to as \textit{LFMDF1} hereinafter.
\end{enumerate}
It is worth noting that the diffusion matrix, $B_{s,ij}$, 
is diagonal but not isotropic. 
This is crucial to reproduce a  correct energy flux from the resolved scales
to the unresolved ones, and represents a necessary requirement to consider the model
acceptable~\cite{minier2014guidelines}.
Using the same closure as that of single-phase flows, namely
$B_{s,ij}= \sqrt{C_0\, \widetilde{\epsilon}}\, \delta_{ij}$, is 
inconsistent with the modeled SGS dissipation rate $\widetilde{\epsilon}$. 

When dealing with dispersed flows, a limit case of particular importance to assess the capability of a
SGS particle model is that of inertia-free particles. These particles behave like fluid tracers
and are characterized by $\tau_p\rightarrow 0$:
The particle model must be consistent with a correct model in this
limit~\cite{minier2014guidelines}. When $\tau_p\rightarrow 0$, our model reduces to:
\begin{align}
 dx_{p,i} &= U_{p,i}\, dt~, \\
 U_{p,i} &= U_{s,i}~,  \\
 dU_{s,i} &= -\frac{1}{\rho_f}\frac{\partial \widetilde{P}}{\partial x_i}\, dt
+\nu_f \Delta \widetilde{U}_i
- \frac{U_{s,i} - \widetilde{U_{i}}}{T_{L,i}}\, dt 
+ \sqrt{C_0 \widetilde{\epsilon}}\, dW_i~,
\label{eq:sde-fluid}
\end{align}
which is the stochastic system equivalent to the Velocity Filtered Density Function (VFDF) model proposed
by Gicquel et al. for the fluid \cite{gicquel2002velocity}.
This model is consistent with the exact zero-th and first moment equations; but more complete models
for the second central moment are also available \cite{gicquel2002velocity,Dre_98,War_04}.

\section{Numerical method}
\label{sec:numerical approach}
The numerical solution of the LES/LFMDF model is obtained
using a hybrid Eulerian mean-field LES/Lagrangian Monte Carlo procedure,
where the filtered fluid properties are computed on a mesh while the statistics
of the dispersed phase are calculated from particles moving in the computational domain.
This procedure has been used previously in the context of RANS \cite{Pei_06}.
Specifically, let $\{{\bf Y^{[x]}}\}$ be the set of filtered fluid flow fields at the
different mesh points and let $\{{\bf Y}^{(N)}\}$ be the set of filtered fluid flow
fields interpolated at particle locations. Let $\{{\bf Z}^{(N)}\}$ be the set of variables
``attached'' to the particles and let $\{{\bf Z^{[x]}}\}$ be the set of statistics
(defined at cell centres) extracted from $\{{\bf Z}^{(N)}\}$. 
The first step (operator $F$) is to solve the PDEs for the fluid:
\begin{equation} 
\{{\bf Y^{[x]}}\}(t_n)
\xrightarrow{F} \{{\bf Y^{[x]}}\}(t_{n+1})~.
\end{equation}
The second step (projection, operator $P$) consists of calculating the filtered
fluid properties and the filtered particle properties at particle locations:
\begin{equation} 
\{{\bf Y^{[x]}}\}(t_n)  \text{\, and \,} \{{\bf Z^{[x]}}\}(t_n)
\xrightarrow{P}
\{{\bf Y}^{(N)}\}(t_n) \text{\, and \,} \{{\bf Z}^{(N)}\}(t_n)~.
\end{equation}
Then, the stochastic differential system can be integrated in time (operator $T$):
\begin{equation} 
\{{\bf Z}^{(N)}\}(t_n)
\xrightarrow{T} \{{\bf Z}^{(N)}\}(t_{n+1})~.
\end{equation}
Finally, from the newly computed (at particle locations) set of variables, new statistical
moments are evaluated at cell centres (operator $A$):
\begin{equation} 
\{{\bf Z}^{(N)}\}(t_{n+1}) \xrightarrow{A} \{{\bf Z^{[x]}}\}(t_{n+1})~.
\end{equation}

The operator $F$ is  a pseudo-spectral method based on trasforming the field variables into
wavenumber space, using Fourier representations for the periodic streamwise and spanwise
(homogeneous) directions and a Chebyshev representation for the wall-normal (non-homogeneous)
direction. A two-levelg, explicit Adams-Bashforth scheme for the nonlinear terms, and an implicit
Crank-Nicolson scheme for the viscous terms are employed for time advancement \cite{Mar_02}.
The projection step, required to evaluate fluid and particle quantities at particle positions, is achieved
with three different techniques:
\begin{itemize}
\item \textit{no-interpolation} (zero-th order, not symmetric in the wall normal direction): The values of
the filtered quantities at the upstream neighbour node of the cell containing the particle are used.
\item \textit{NGP} (Nearest Grid Point, symmetric in the wall-normal direction): The average
values of the filtered quantities at each node of the cell containing the particle are used. 
\item \textit{interpolation}: A second-order interpolation of the Eulerian quantities at grid nodes is
performed to obtain quantities at the particle position. 
\end{itemize}
Previous studies have shown that no improvement is obtained using
higher-order interpolation schemes \cite{Pei_06} .
In fact, higher-order schemes may even lead to larger errors in hybrid formulations like the one
considered here.

The local instantaneous properties of the dispersed phase are obtained by solving the set of SDEs via
the operator $T$.  In particular, the numerical solution of the modelled stochastic equations is obtained
representing the modelled LFMDF through an ensemble of $N$ statistically identical Monte Carlo particles. 
Each of these particles carries information pertaining to the fluid velocity seen by the particle,
$\mathbf{U}_s^{(n)}(t)$, to the particle velocity, $\mathbf{U}_p^{(n)}(t)$, and to the particle position,
$\mathbf{x}_p^{(n)}(t)$, where $n=1,2,\ldots,N$.
This information is updated upon time-integration
of Eqns. (\ref{eq:sdeXp})-(\ref{eq:sde}).
This system of SDEs has multiple scales and may become stiff, in particular for particles with very
small inertia. Moreover, in wall-bounded flows the characteristic fluid time scales become smaller in the
near-wall region, thus complicating the integration.
For these reasons, an ad-hoc unconditionally-stable, second-order accurate
numerical scheme has been developed and implemented here. The scheme is based on that put forward in the RANS context \cite{Pei_06}: It adopts the It\^o's convention and is developed starting from the analytical solution of Eqns. (\ref{eq:sdeXp})-(\ref{eq:sde}) with constant coefficients.
Such a scheme ensures stability and consistency with all limit cases.
The first-order  scheme is the following Euler-Maruyama:
\begin{eqnarray} \label{eq:first_order_scheme1}
  x_{p,i}^{n+1} &=& x_{p,i}^n + A_1\,U_{p,i}^n + B_1\,U_{s,i}^n
  + C_1\,[T_i^n C_i^n] + \Omega _i^n~,\\
   \label{eq:first_order_scheme2}
  U_{p,i}^{n+1} &=& U_{p,i}^n\, \exp(-\Delta t/\tau_p^n)
+ D_1\,U_{s,i}^n + [T_i^n C_i^n](E_1-D_1)
+ \Gamma _i^n~,\\
 \label{eq:first_order_scheme3}
  U_{s,i}^{n+1} &=& U_{s,i}^n\, \exp(-\Delta t/T_i^n)
+ [T_i^n C_i^n] [1-\exp(-\Delta t/T_i^n)]
+ \gamma _i^n~, 
\end{eqnarray}
where the coefficients are given by the following relations:          
\begin{align}
& \quad A_1 = \tau_p^n\,[1-\exp(-\Delta t/\tau_p^n)]~,\notag\\
& \quad B_1 = \theta_i ^n\,[T_i^n(1-\exp(-\Delta t/T_i^n)-A_1]
      \quad \text{with}\quad \theta_i ^n = T_i^n/(T_i^n-\tau_p^n)~,\notag\\
& \quad C_1 = \Delta t - A_1 - B_1~, \notag \\
& \quad D_1 = \theta_i ^n [\exp(-\Delta t/T_i^n)-\exp(-\Delta
  t/\tau_p^n)]~,\notag\\ 
& \quad E_1 = 1 - \exp(-\Delta t/\tau_p^n)~.\notag  
\end{align}  
and $\gamma_i^n,\Gamma_i^n,\Omega_i^n$ are stochastic integrals.
The details of the scheme as well as the analytical solutions are given in Appendix \ref{app:scheme}.
The second-order scheme is derived using a predictor-corrector technique, in which
the prediction step is the first-order scheme given by
Eqns. (\ref{eq:first_order_scheme1})-(\ref{eq:first_order_scheme3}) \cite{Pei_06}.

Particle statistics are evaluated by considering the ensemble of particles $N_{E}$ located within a
small volume of fluid $\delta V$ (a box of size $\Delta_{E,1} \times \Delta_{E,2} \times \Delta_{E,3}$)
centered around a given point $\bm x$. This ensemble provides one-time one-point statistics.
For reliable statistics with minimal numerical dispersion, it is desirable to minimize the size of
the averaging domain, namely $\Delta_{E}=\sqrt[3]{\Delta_{E,1} \Delta_{E,2} \Delta_{E,3}} \to 0$,
and maximize the number of Monte Carlo particles, namely $N_{E} \to \infty$. By doing so, the ensemble
statistics tend to the desired filtered values:
\begin{equation}
\left.\begin{split}
\widetilde{ a}_E &= \frac{1}{N_E} \sum_{n \in \Delta_E} a^{(n)} \xrightarrow[\Delta_E \to 0]{N_E \to \infty} \widetilde{a} \\
\tau_E(a,b) &= \frac{1}{N_E} \sum_{n \in \Delta_E} (a^{(n)} - \widetilde{a}_E )(b^{(n)} - \widetilde{b}_E ) \xrightarrow[\Delta_E \to 0]{N_E \to \infty} \tau(a,b)
\end{split}\right.
\label{eq:averaging}
\end{equation}
where $a^{(n)}$ and $b^{(n)}$ denote typical information 
 carried by the \textit{n}-th particle, for instance its velocity components.
Since we are adopting a Monte Carlo procedure in a LES/LFMDF approach, the quantities
obtained following Eqn. (\ref{eq:averaging}) are filtered Eulerian quantities, $\widetilde{ a}$, and subgrid quantities,
$\tau(a,b)$, respectively.
For example, one can evaluate the particle filtered velocity as:
\begin{equation}
\widetilde{U}_{p,i} (\bm x)  \simeq \frac{1}{N_x} \sum_{n=1}^{N_x} U_{p,i}^{(n)}~.
\label{eq:lagran-filtering}
\end{equation}
Analogous expressions can be written for all other filtered quantities. 


The mean-field LES solver also computes the filtered fluid velocity field so that there is a ``redundancy" of
the first filtered moments in the $\tau_p\rightarrow 0$ limit. In this case, both the
spectral method and the Monte Carlo procedure yield the solution for the particle number density and velocity fields. These fields are referred to as
``duplicate fields" hereinafter, and can be exploited to assess the accuracy
of the model~\cite{Muradoglu:1999p4184,Jenny:2001p4207}.
The characteristics of our scheme are summarized in Table \ref{table:duplicate}.
\begin{table}[b]
\begin{center}
\begin{tabular}{ c| c| c| c}
\hline
 	 spectral LES  & Particle solver  & mean-field  & duplicate   \\
 	  variables &  variables &  variables &  fields  \\
 	   &   &   &  (fluid limit) \\
\hline
 $\widetilde{U}_i$ & 
 $X_{p,i}$ 
 & $\widetilde{U}_i~, ~\frac{\partial \widetilde{P}}{\partial x_i}\,$  
 &  $\rho_f~$   \\
 $\widetilde{P}$ & 
 $U_{p,i}~,~ U_{s,i}$&
 $ \frac{\partial \widetilde{U}_i}{\partial x_j}\,~\Delta \widetilde{U}_i~$ &
 ~$\widetilde{U}_i$ \\
\hline
\end{tabular}
\caption{Summary of the LES/LFMDF solution procedure.}
\label{table:duplicate}
\end{center}
\end{table}

\section{Results}

In the present study, the LES/LMFDF approach is applied to track inertial particles in
gas-solid turbulent channel flow. The fluid considered is air (assumed to be incompressible
and Newtonian) with density $\rho_f = 1.3 \; kg/m^3$ and kinematic viscosity
$\nu_f = 1.57 \cdot 10^{-5} \; m^2/s$. \\
The reference geometry consists of two infinite flat parallel walls: the origin of the coordinate system is located at the center of the channel,
with the $x-$, $y-$ and $z-$ axes pointing in the streamwise, spanwise and wall-normal directions, respectively. Periodic boundary conditions
are imposed on the fluid velocity field in $x$ and $y$, and no-slip boundary conditions are imposed at the walls. Calculations were performed
on a computational domain of size $4 \pi h \times 2 \pi h \times 2h$ in $x$, $y$ and $z$, respectively \cite{Sol_09}. The domain was discretised using
a $32 \times 32 \times 33$ grid with uniform cell spacing in the homogeneous directions and non-uniform
cell distribution in the wall-normal direction (Chebyshev collocation points) \cite{Mar_08a}.
Simulations were performed with a coarsening factor $CF=8$ with respect to the reference DNS, at a shear Reynolds number $Re_{\tau}=300$ based on the half width $h$ of the channel, and using a fixed time step
(see Table \ref{tab:simulation-param}). 
Particles with density $\rho_p = 10^3~kg / m^3$ and Stokes numbers as given in Table \ref{tab:particle-param},
were injected in the flow at randomly-chosen locations under fully-developed flow conditions. 
Since we are concerned with a Monte-Carlo simulation, a large number of particles is required to minimize statistical errors. In the consistency assessments (see Section \ref{sec:consist}), the number of particles
per cell was varied selecting $N_{pc}=20, 40$, and $80$, while simulations with inertial particles were performed
imposing $N_{pc}=40$: This latter value corresponds to a total number of particles $N \simeq 1.31 \cdot 10^6$ in the domain.

In the following, both instantaneous and time-averaged results are discussed.
In particular, we examine Reynolds averaged statistics, denoted by an overbar and obtained upon
averaging the filtered velocity over the homogeneous flow directions and in time. 
\begin{table}[]
\begin{center}
\begin{tabular}{c c c c}
\hline
Time step & $\Delta t$ & $4.2 \cdot 10^{-5}$ & $[s]$ \\
& $\Delta t^+= \Delta t u_{\tau}^2/\nu_f$ & $0.15$ & $[w.u.]$ \\
DNS grid size & $N_x \times N_y \times N_z$ & $256 \times 256 \times 257$ & \\
LES grid size & $N_x \times N_y \times N_z$ & $32 \times 32 \times 33$ & \\
\hline
\end{tabular}
\end{center}
\caption{Simulation parameters for the fluid. Superscript + represents variables in wall units, obtained using
the shear velocity and the fluid kinematic viscosity.}
\label{tab:simulation-param}
\end{table}
%
\begin{table}[b]
\begin{center}
\begin{tabular}{c c c c}
\hline
$St$ & $\tau_p \; [s]$ & $d_p^+ \; [w.u.]$ & $d_p \; [\mu m]$ \\
\hline
$1$ & $0.283 \cdot 10^{-3}$ & $0.153$ & $10.2$ \\
$5$ & $1.415 \cdot 10^{-3}$ & $0.342$ & $22.8$ \\
$25$ & $7.077 \cdot 10^{-3}$ & $0.763$ & $50.9$ \\
\hline
\end{tabular}
\end{center}
\caption{Simulation parameters for the particles.}
\label{tab:particle-param}
\end{table}

\subsection{Assessment of consistency and convergence} \label{sec:consist}
The purpose of this section is to demonstrate 
the consistency of the LFMDF formulation in the $\tau_p \to 0$ limit,
and to show its convergence.
To these objectives, the results obtained via the LES/mean-field are compared against
those provided by the LFMDF approach. Given the accuracy of the spectral method,
such a comparative validation represents a robust 
way to assess the performance of the LFMDF solution provided by the Monte Carlo simulation.
We are particularly interested in examining the particle velocity statistics, but also the particle number density distribution, which is the
macroscopic result of turbophoresis \cite{Mar_02,Sol_09} and should remain uniform
in the whole domain when $\tau_p \to 0$. 
For these observables, we compare the statistics obtained from the Monte Carlo simulation,
namely from the solution of Eqns. (\ref{eq:sdeXp})-(\ref{eq:sde}), with those of the
Eulerian pseudo-spectral simulation, which solves for Eqns. (\ref{fluid: exact Cont})-(\ref{fluid: exact NS}).
As mentioned, in the fluid limit this is equivalent to solving Eq. (\ref{eq:sde-fluid}), and the resulting
duplicate fields (indicated in Table \ref{table:duplicate}) should be consistent. 
The values suggested in the literature for the model parameters are chosen here:
$C_0=2.1~,~C_{\epsilon}=1$, $\beta = 0.8$ \cite{Min_01}.
We have also checked the convergence with respect to $N_{pc}$, which is achieved for
$N_{pc} \ge 40$.

\begin{figure}[b]
{\includegraphics[width=.42\textwidth]{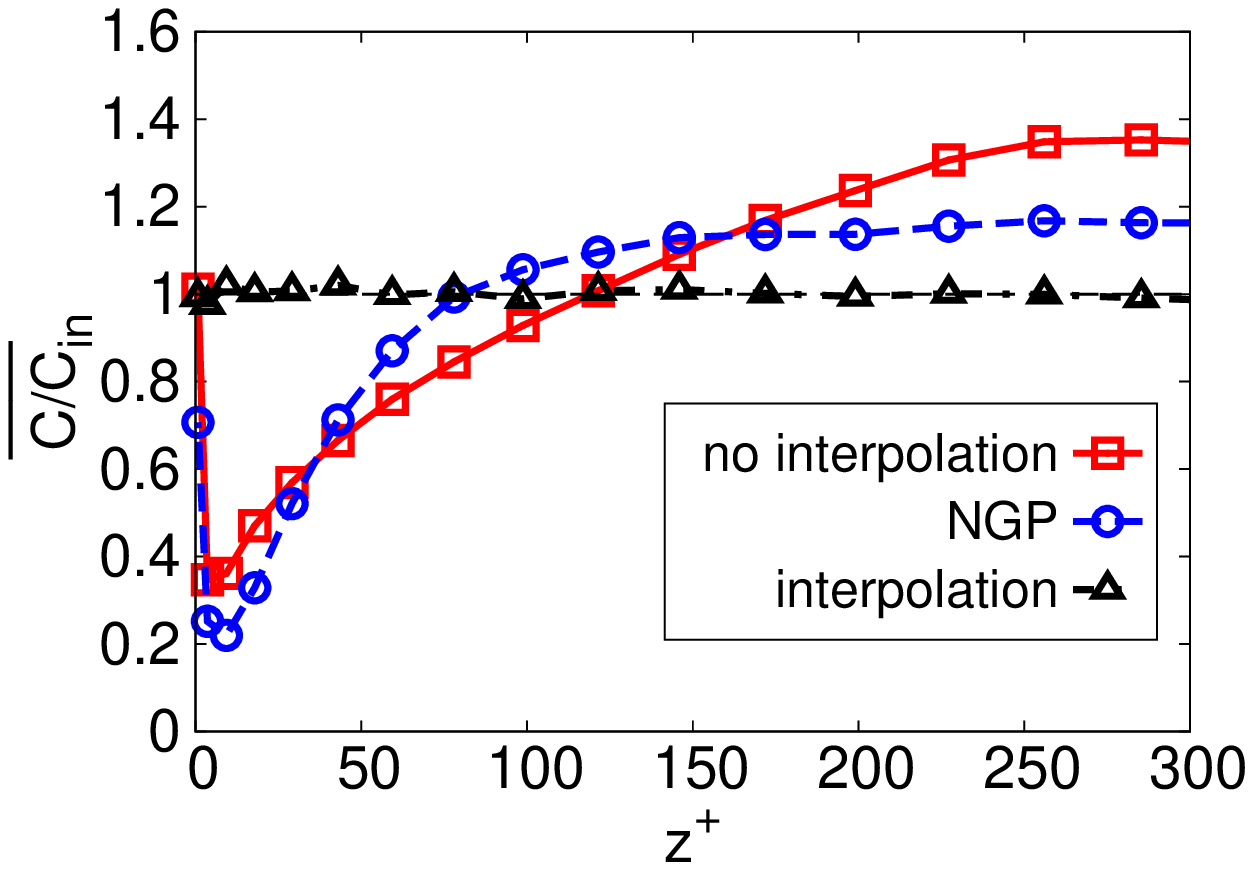}}
{\includegraphics[width=.42\textwidth]{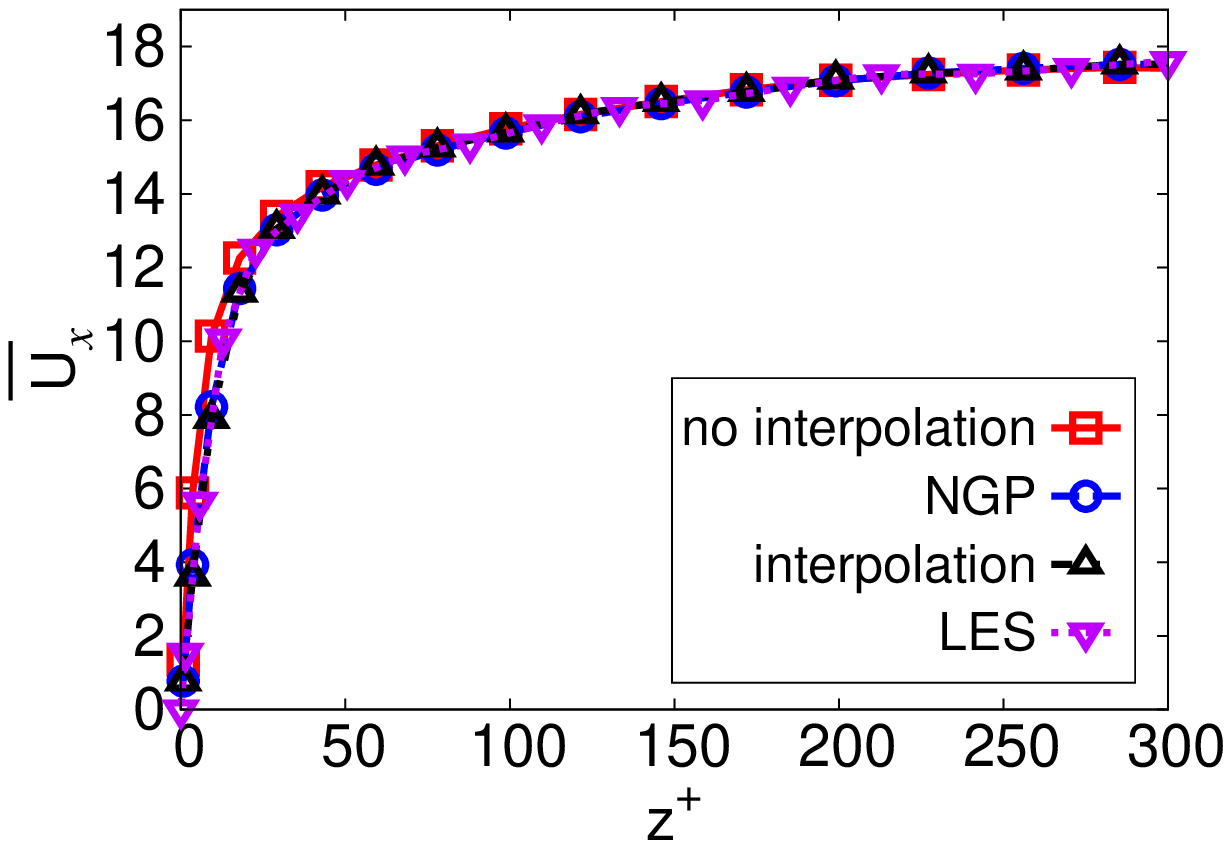}}
\caption{Reynolds-averaged particle number density (a) and filtered streamwise velocity (b),
obtained with different SGS particle models: no-interpolation of LES and particle quantities ({\color{red}{$\square$}}), NGP
interpolation ({\bf {\color{blue}{$\circ$}}}) and second-order interpolation ($\triangle$).
Downward triangles ({\color{magenta}{$\triangledown$}}) in panel (b) refer to the filtered streamwise velocity provided by LES. The time window for averaging is $\Delta t^+ = 3000$,
in wall units.}
\vspace{-8.8cm}
\hspace{-3.7cm} (a) \hspace{6.4cm} (b)
\vspace{8.8cm}
\label{Fig:varying-interp}
\end{figure}
\begin{figure}[h]
{\includegraphics[width=.42\textwidth]{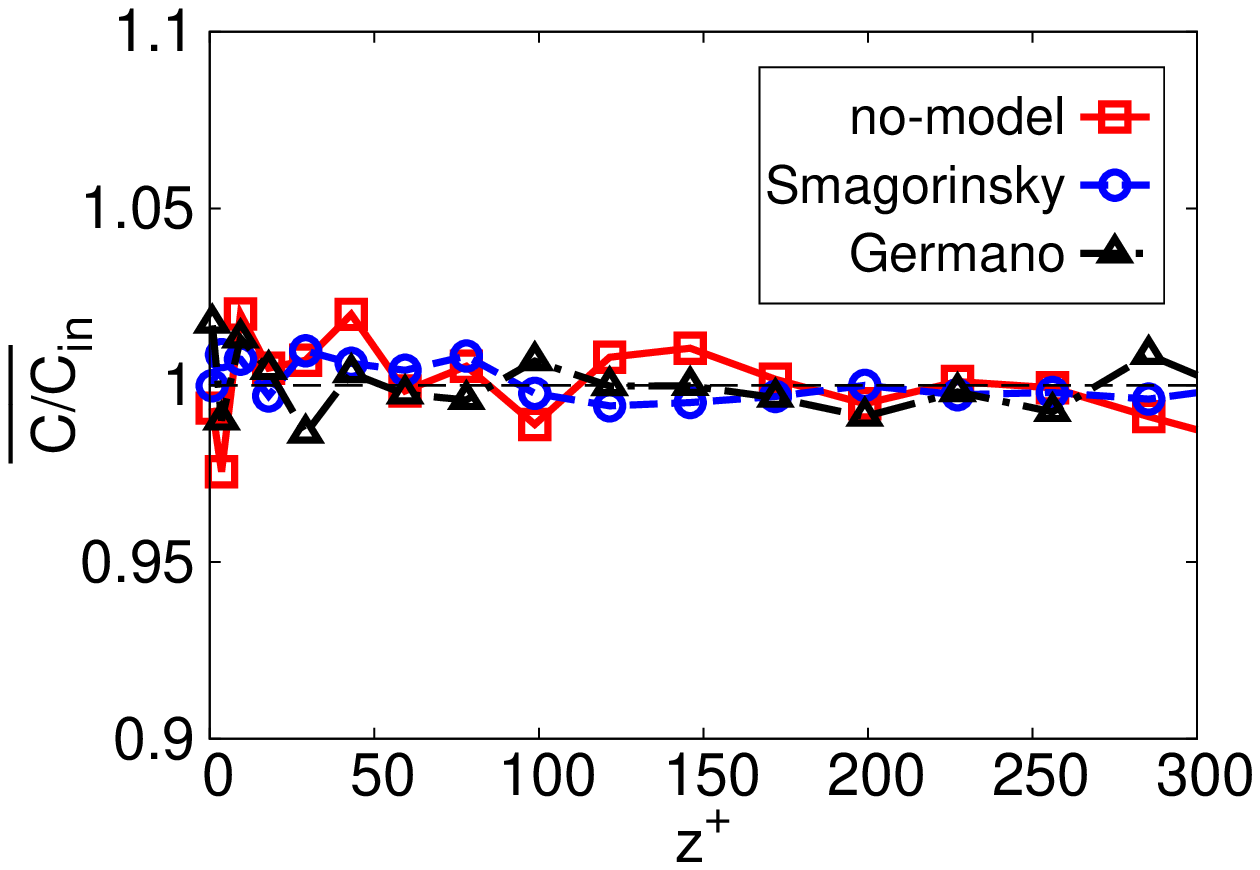}}
{\includegraphics[width=.42\textwidth]{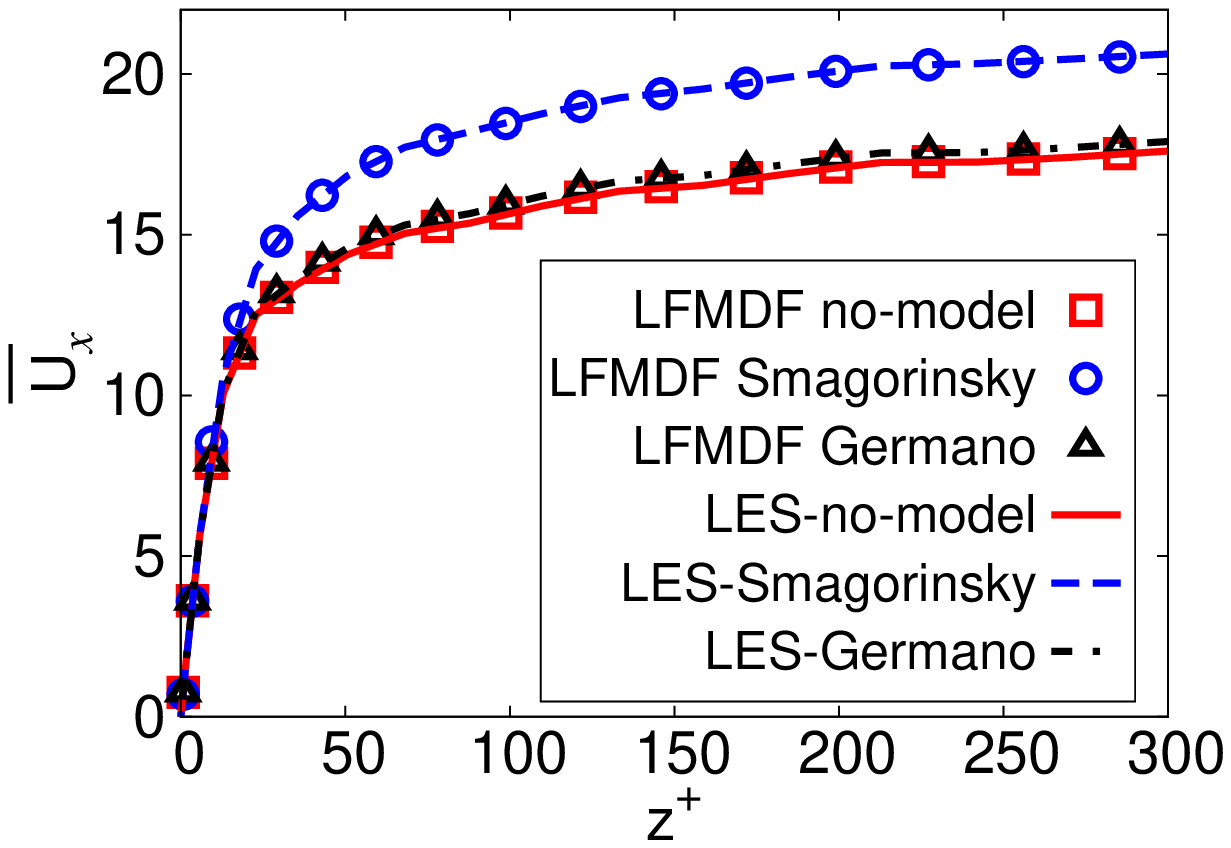}}
\caption{Reynolds-averaged particle number density (a) and filtered streamwise velocity (b),
obtained with different SGS models for the fluid: No-model ({\color{red}{$\square$}}, {\color{red}{$-$}}), Smagorinsky model ({\color{blue}{$\circ$}}, {\color{blue}{$- -$}}) and
Germano dynamic model ({\color{black}{$\triangle$}}, {\color{black}{$- \cdot$}}). 
The time window for averaging is $\Delta t^+ = 3000$,
in wall units.}
\vspace{-8.cm}
\hspace{-3.6cm} (a) \hspace{6.4cm} (b)
\vspace{7.3cm}
\label{Fig:variable-fluid-model}
\end{figure}

\begin{figure}[h]
\vspace{0.3cm}
{\includegraphics[width=.42\textwidth]{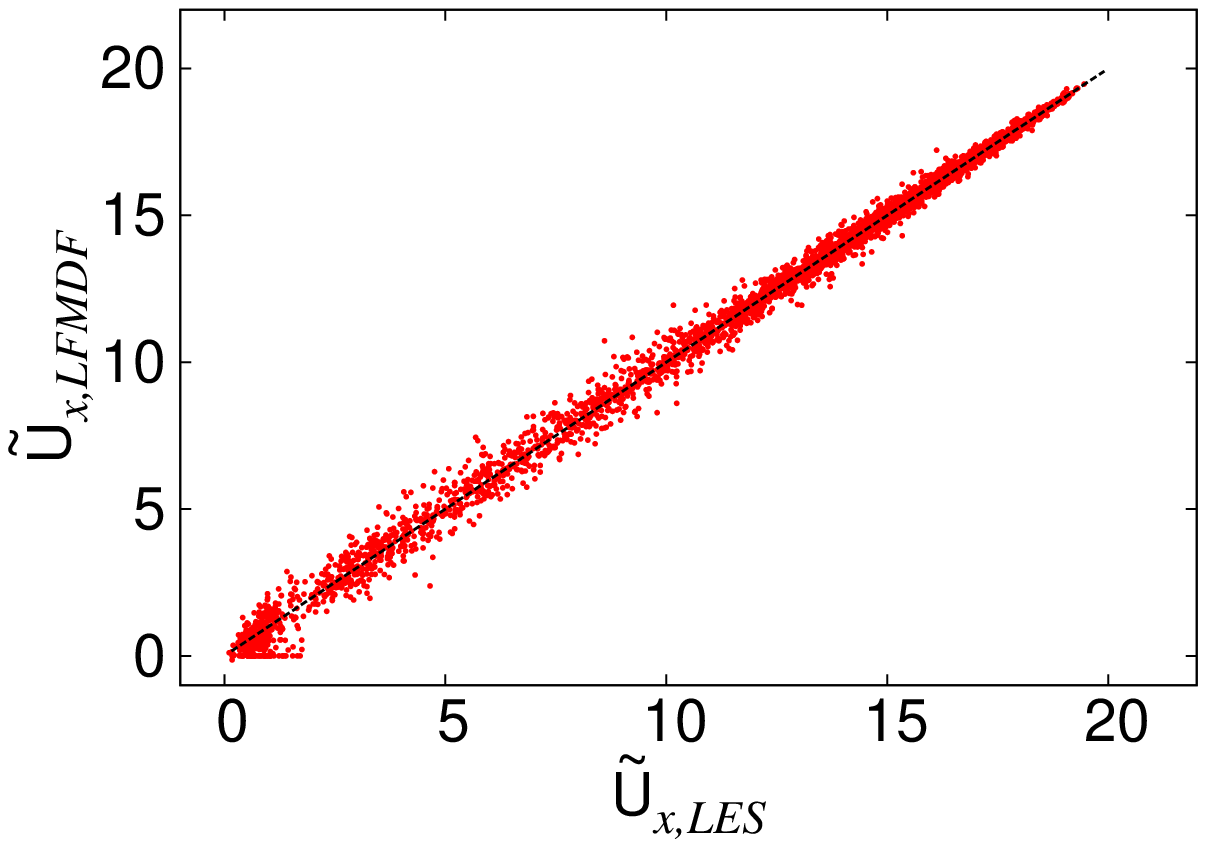}}
{\includegraphics[width=.42\textwidth]{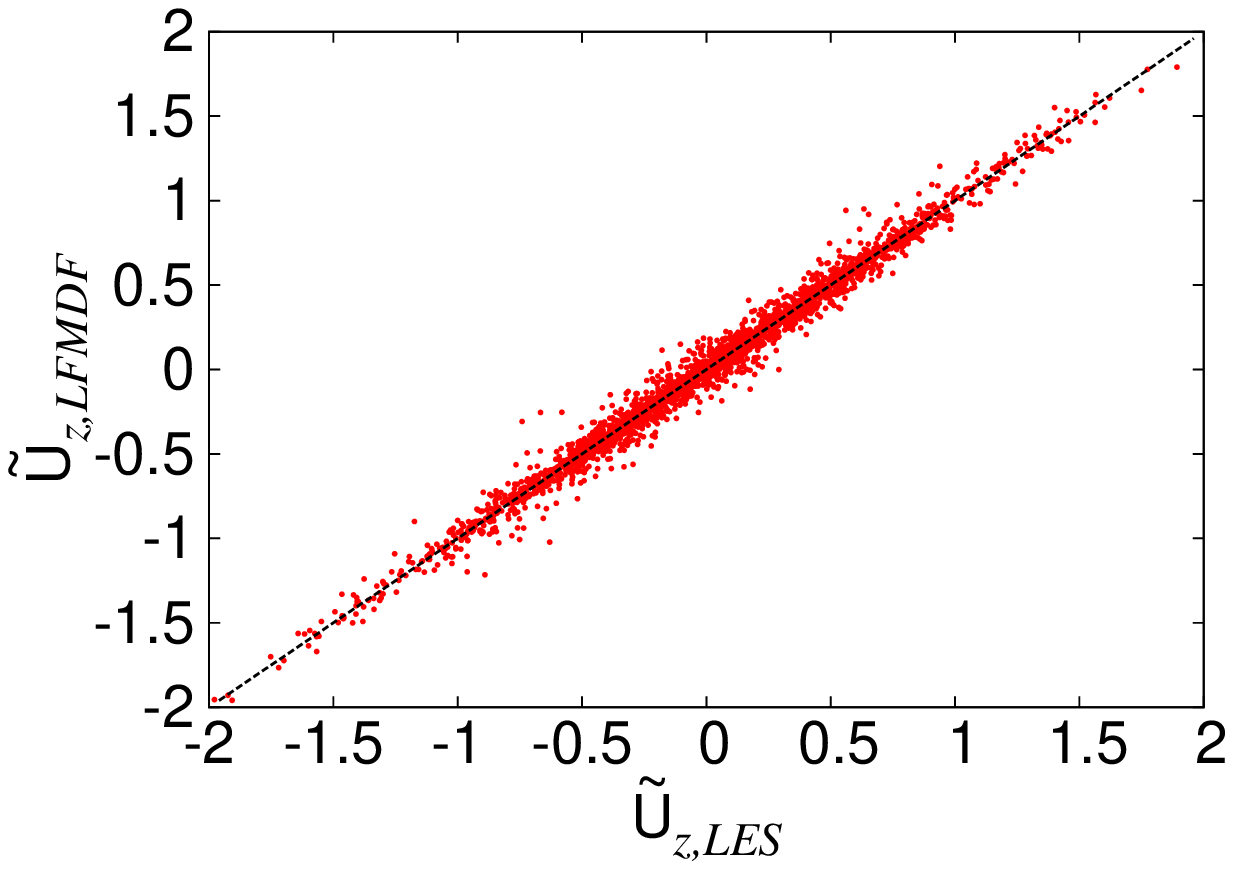}}
\caption{Scatter plot correlating particle velocity components evaluated from LES and from LFMDF:
(a) streamwise component, (b) wall-normal component.}
\vspace{-6.5cm}
\hspace{-3.4cm} (a) \hspace{6.5cm} (b)
\vspace{6.5cm}
\label{Fig:correlation}
\end{figure}

Figure \ref{Fig:varying-interp} shows the Reynolds-averaged particle number density, $\overline{C / C_{in}}$
(with $C_{in}$ the number density at the time of particle injection),
and particle streamwise velocity, ${\overline{ U_x}}$ along the wall-normal coordinate. The different profiles correspond to different interpolation techniques.
To avoid cross-effects, no subgrid model is used in the Eulerian simulation.
While velocity appears unaffected by the particular interpolation technique employed (results are
perfectly consistent), particle number density is sensitive. In particular significant errors in the near-wall
region are found when no interpolation is performed or when the nearest-grid-point technique is
used. A second-order interpolation, however, is sufficient to recover the expected behaviour and
ensure $\overline{C / C_{in}} \simeq 1$ everywhere (as expected for tracer particles). 
In figure \ref{Fig:variable-fluid-model}, the averaged number density profile and the averaged velocity provided
by the different SGS models for the fluid are shown. The LFMDF model appears to be consistent
with all models tested, since the $\overline{C / C_{in}}$ profile remains uniform once the
stationary state is reached and the velocity is (again) perfectly consistent. It is also observed that,
in the $\tau_p \to 0$ limit, the first moments of the Germano model are nearly the same as those
obtained without SGS model.
Therefore results discussed hereinafter refer to simulations performed using the Germano
model for the fluid phase, unless otherwise stated.
A further proof of consistency is provided by figure \ref{Fig:correlation}, which shows
the scatter plots of the streamwise
and wall-normal velocity components, indicated as $\widetilde{U}_x$ and $\widetilde{U}_z$
respectively.
Velocities in the Eulerian simulations were evaluated at the center of the computational cells.
The velocity correlation is quite satisfactory, except perhaps for very small values of $\widetilde{U}_x$.

\begin{figure}
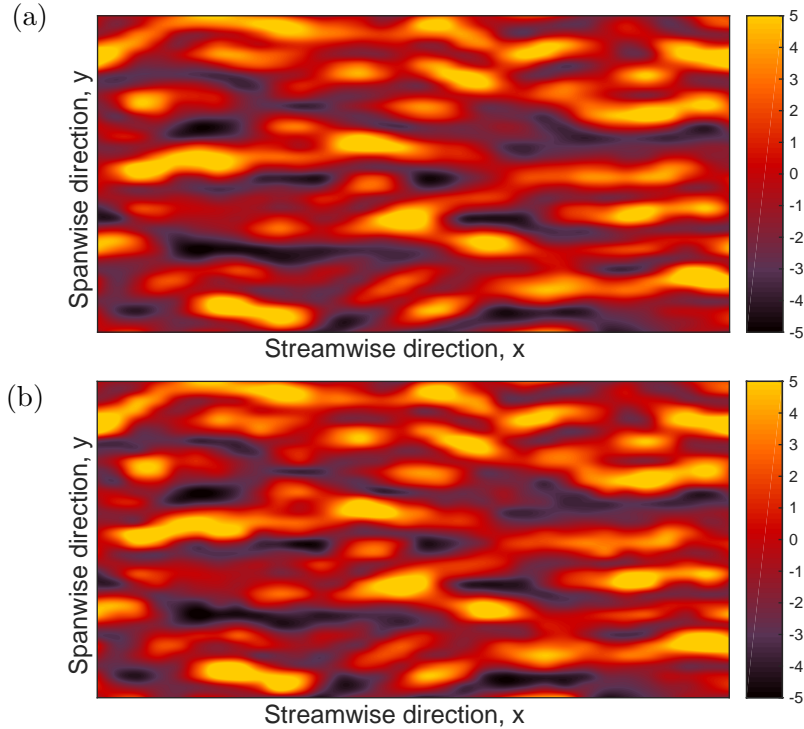

\centering
\hspace{-0.25cm}
{\includegraphics[width=0.6\textwidth]{figure4a.eps}}
\vspace{1cm}
{\includegraphics[width=0.6\textwidth]{figure4b.eps}}
\vspace{-0.5cm}
\caption{Near-wall fluid velocity streaks. Low-speed (high-speed) streaks are rendered using
colored contours of negative (positive) streamwise fluctuating velocity on a horizontal plane at
$z^+=10$ from the wall. Panel (a) refers to the Eulerian LES, performed with no SGS model for the particles
and with $N_{pc}=40$; panel (b) refers to the LFMDF simulation.
}
\vspace{-13.8cm}
\hspace{-11.0cm}
(a)

\vspace{4.3cm}
\hspace{-11.0cm}
(b)
\vspace{8.5cm}
\label{Fig:color}
\end{figure}

To assess the consistency of the LFMDF formulation from a physical (and more intuitive) point of view,
in Fig. \ref{Fig:color}, we compare the near-wall fluid streaks that can be rendered from the Eulerian LES
(panel a) and from the Monte Carlo LFMDF simulation (panel b).
Streaks are known to play a crucial role in determining the transport mechanisms in turbulent boundary
layer \cite{Pic_05,Mar_02}, and are visualised here by instantaneous contour plots of the fluctuating streamwise
velocity on a $x$-$y$ plane located at a distance $z^+=10$ from the wall.
Visual inspection shows only small differences in the color map, indicating that the streaks, and
indirectly the near-wall turbulent coherent structures that generate it, are indeed recovered by the
LFMDF simulation in the fluid limit.

\begin{figure}[b]
{\includegraphics[width=.55\textwidth]{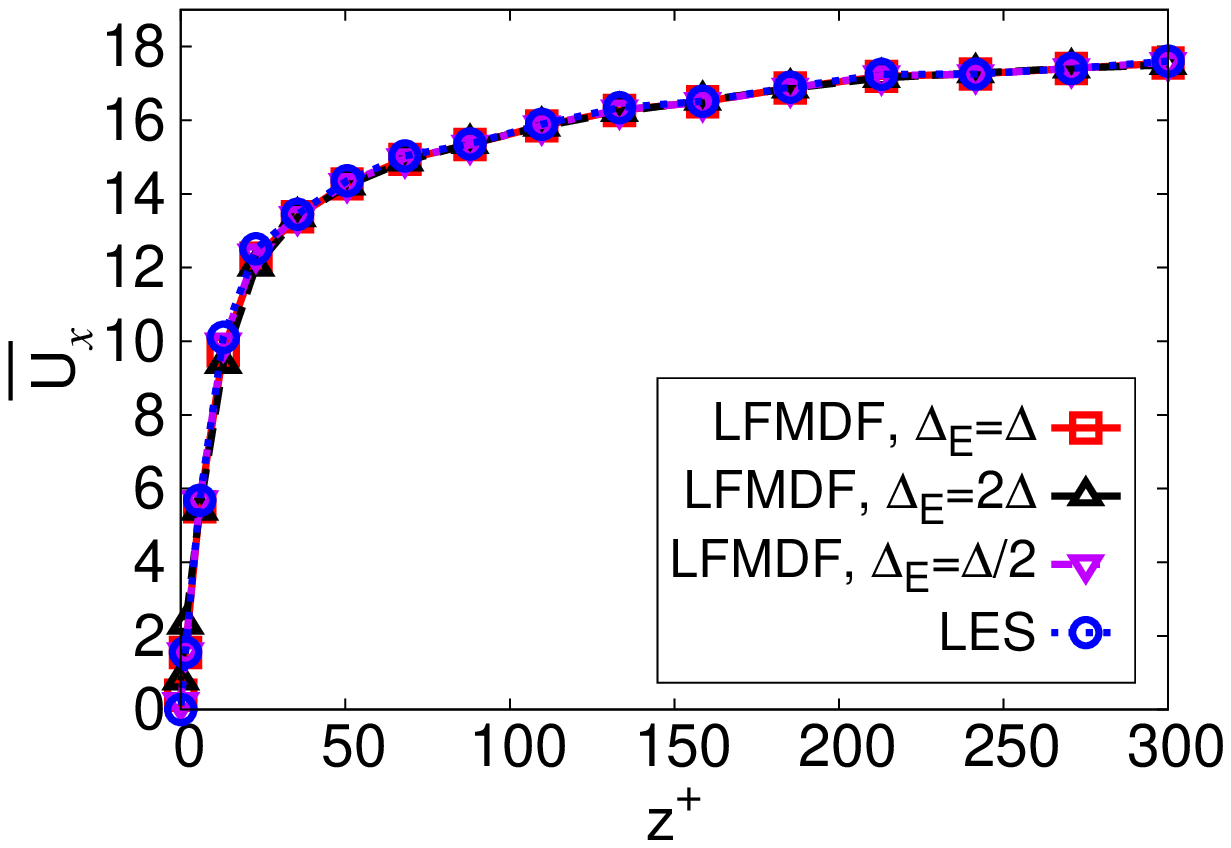}}
\caption{Mean velocity of the filtered streamwise velocity at varying $\Delta_E$. Time window for averaging is $\Delta t^+ = 3000$ with $N_{pc}=40$ particle per cell.}
\label{Fig:variable-filter}
\end{figure}

\begin{figure}[h]
{\includegraphics[width=.55\textwidth]{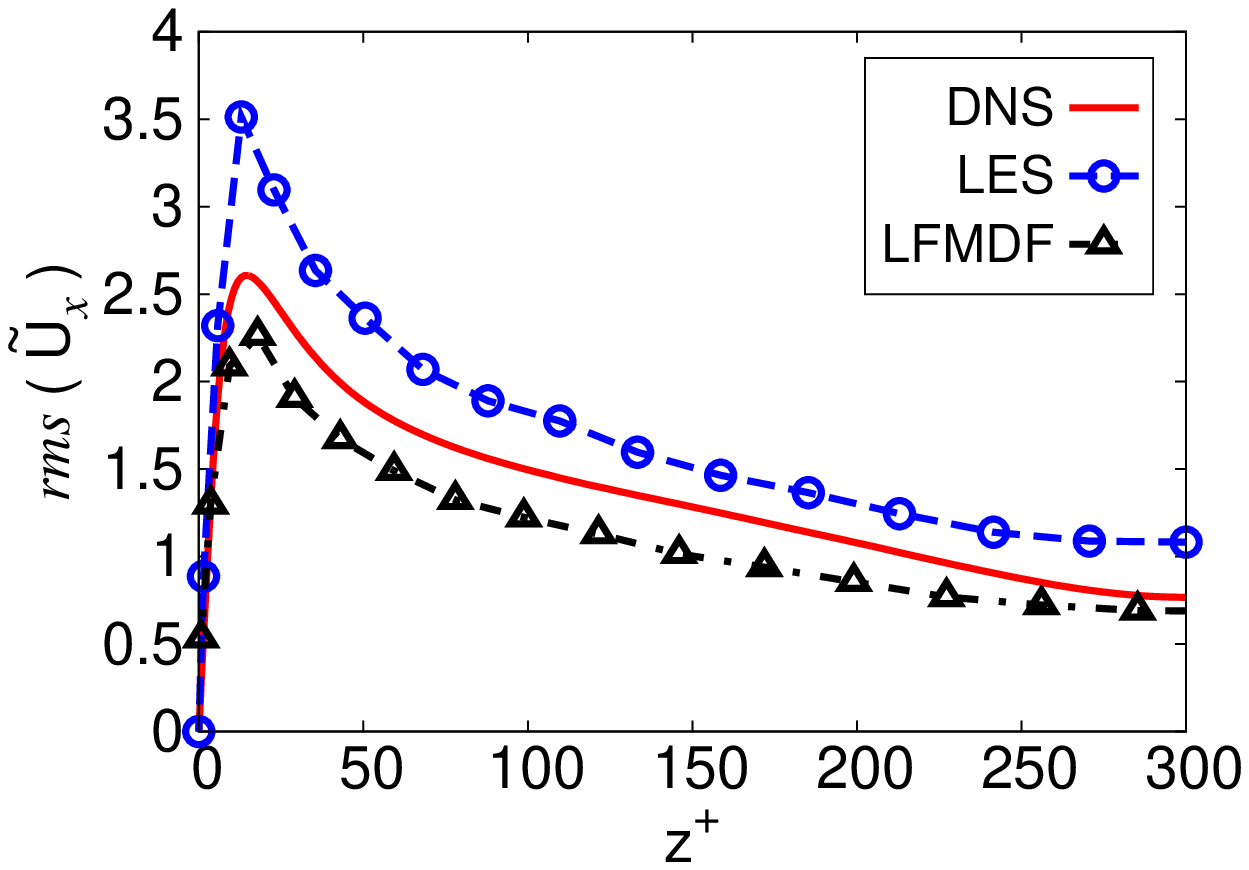}}
\caption{Root mean square of the filtered streamwise velocity.}
\label{Fig:variable-filter2}
\end{figure}

To complete the model assessment, we have also checked the sensitivity of Reynolds averaging
to the size of the reference volume $\delta V$ (introduced in Section IV) over which averaging is performed.
To this aim, we considered different grids made of cubic volumes centered around the LES (Eulerian)
nodes. The size of each volume, $\Delta_E$, was varied to be either smaller or
larger than the cell size $\Delta$ in the reference $32^3$ LES grid.
Figure \ref{Fig:variable-filter} shows the averaged filtered streamwise velocity at varying
$\Delta_E$ (with a fixed number of particles per cell, $N_{pc}=40$).
It can be seen that all profiles overlap even for large $\Delta_E$ ($\Delta_E=2 \Delta$) indicating
that the mean filtered velocity is not sensitive to the size of the averaging volume, at least in the range
of $\Delta_E$ analysed.
For this reason we have chosen $\Delta_E=\Delta$ for all 
simulations.
To test this choice we have also considered higher-order moments,
namely the root mean square (rms) of filtered velocity, and
we have analysed the convergence in relation to the DNS results.
In figure \ref{Fig:variable-filter2} we show the rms of the filtered velocity, defined as
$ rms(\tilde{U})  = \sqrt{(\overline{ \tilde{U} - \overline{\tilde{U}})^2}}$.
The different profiles do not collapse 
and the LFMDF is in better agreement than LES
with DNS,  when the volume size is $\Delta_E=\Delta$,
confirming the validity of the overall method in the fluid limit.
It is worth noting that the discrepancy between Eulerian LES and LFMDF is not related to some incongruity, since
these two models are not fully consistent at the Reynolds-stress level.
As suggested in previous studies \cite{gicquel2002velocity},
an even better convergence to DNS would be probably possible with smaller $\Delta_E$ and much higher $N$.
However, this choice would increase the computational cost considerably thus making the model not relevant
application-wise.
%
%
%
\subsection{Model assessment with inertial particles}
%
%
In this section we validate the LFMDF approach for the case of inertial particles via comparative assessment
against DNS data. In particular, first we exploit DNS to determine the range of
empirical constants appearing in the LFMDF sub-model (\emph{a priori} assessment).
Second, we compare the predictions of the LFMDF-based simulations with the statistics
provided by DNS, which is regarded here as the reference numerical experiment
(\emph{a posteriori} assessment).
In the latter case, comparison is also made with the statistics provided by LES when no
particle SGS model is used, in order to point out the impact of the proposed stochastic
model on statistics.
As mentioned, one of the main difficulties of modelling inertial particle dynamics in LES
is to capture preferential concentration \cite{Mar_08ACME,Mar_08}. Hence, the primary observable
considered for comparative assessment is the \emph{instantaneous} particle number
density distribution along the wall-normal direction. Such comparison is particularly severe
since any error associated with the proposed particle SGS model will inevitably sum up
over time and may thus lead to significant deviations in the final density distribution (we
remark here that all LES/LFMDF simulations are carried out with a rather large coarsening
factor, $CF=8$ with respect to DNS).

\begin{figure}[hb]
{\includegraphics[width=.32\textwidth]{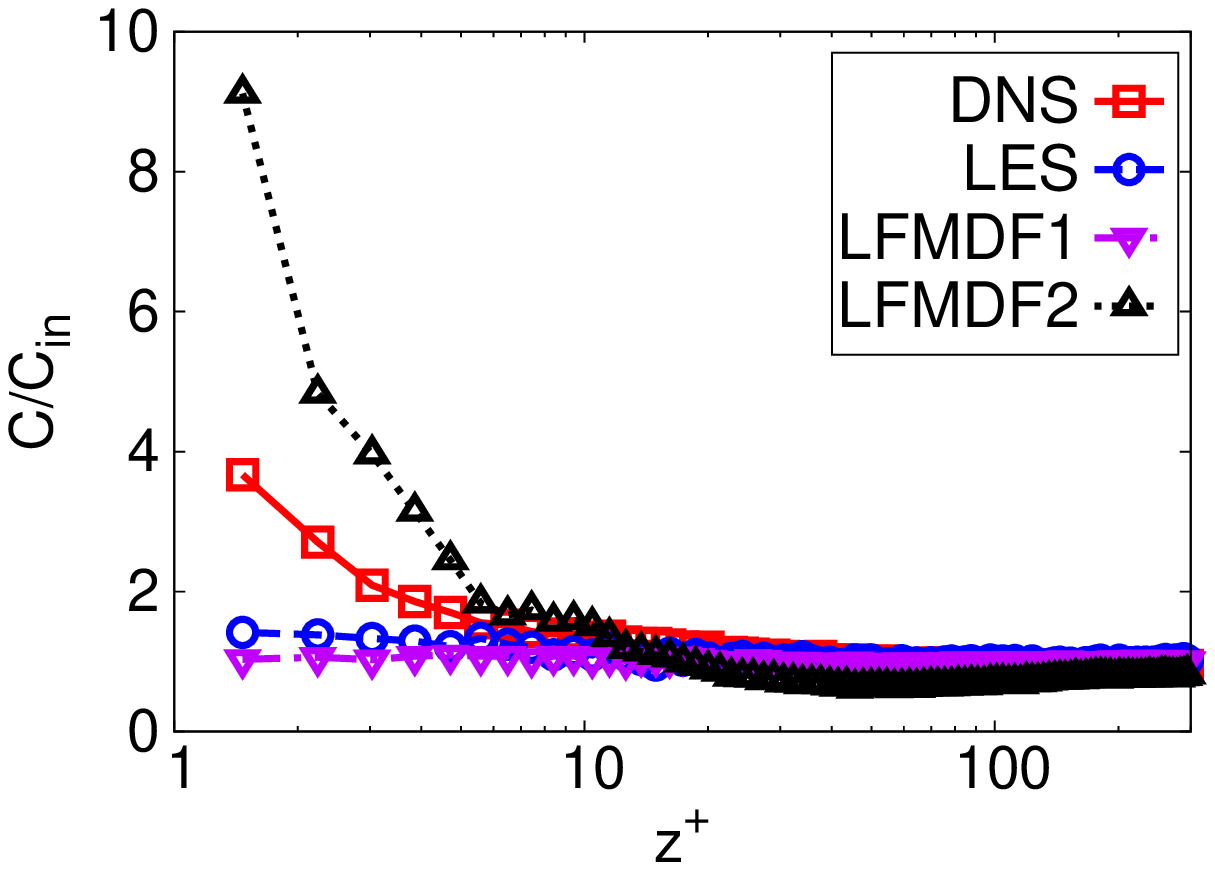}}
{\includegraphics[width=.307\textwidth]{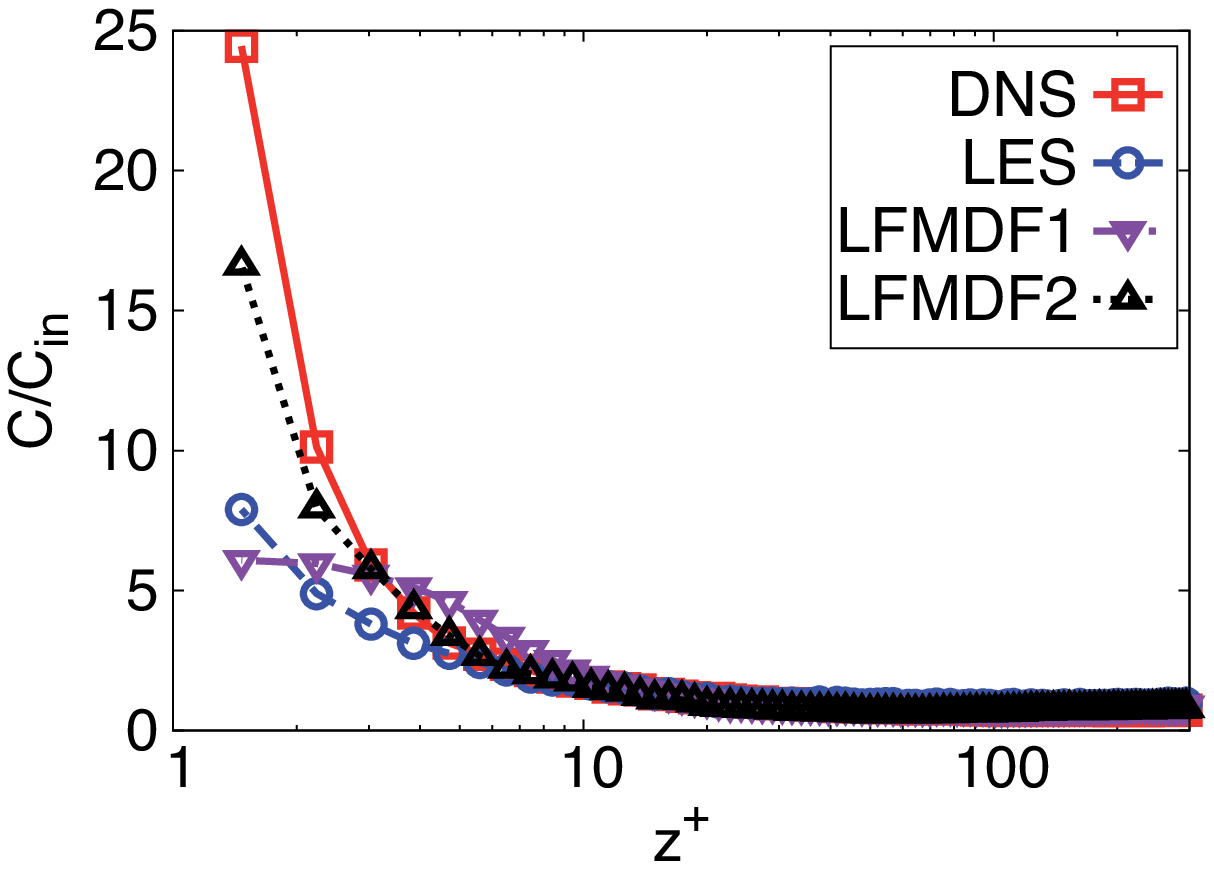}}
{\includegraphics[width=.32\textwidth]{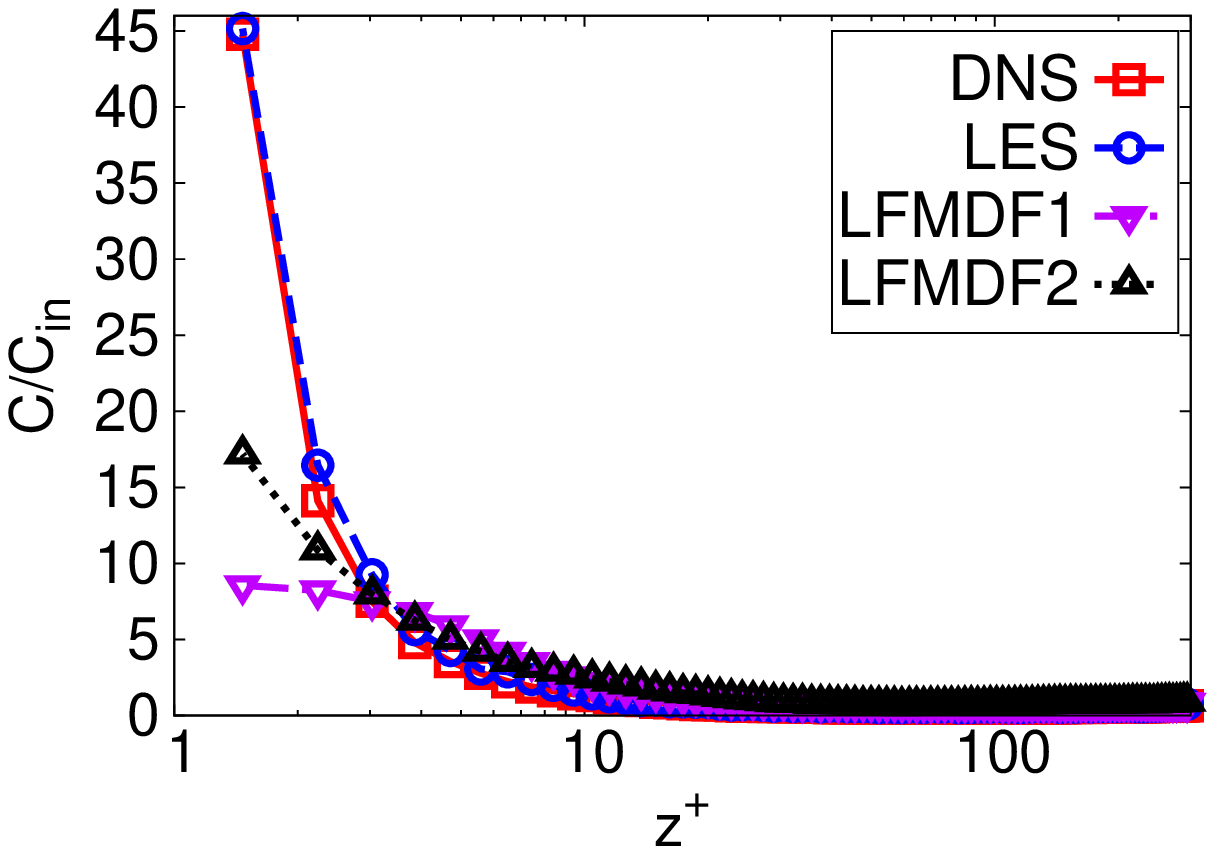}}
\caption{Comparison of particle number density distributions predicted by two different particle SGS model
formulations: Simplified stochastic model (\textit{LFMDF1}, {\color{purple}{$\triangledown$}}) and
complete stochastic model (\textit{LFMDF2}, {\color{black}{$\triangle$}}).
See also Sec. \ref{sec:equiv_stoc_syst}.
Other symbols:
{\color{red}{$\square$}} DNS,
{\color{blue}{$\circ$}} LES without particle SGS model.
Panels: (a) $St=1$ particles, (b) $St=5$ particles, (c) $St=25$ particles.
Profiles are computed at $t^+ =2130$ after particle injection into the flow.}
\vspace{-7.7cm}
\hspace{-1.5cm} (a) \hspace{4.7cm} (b) \hspace{4.7cm} (c)
\vspace{7.8cm}
\label{Fig:concentration-nok}
\end{figure}

\begin{figure}[]
{\includegraphics[width=.32\textwidth]{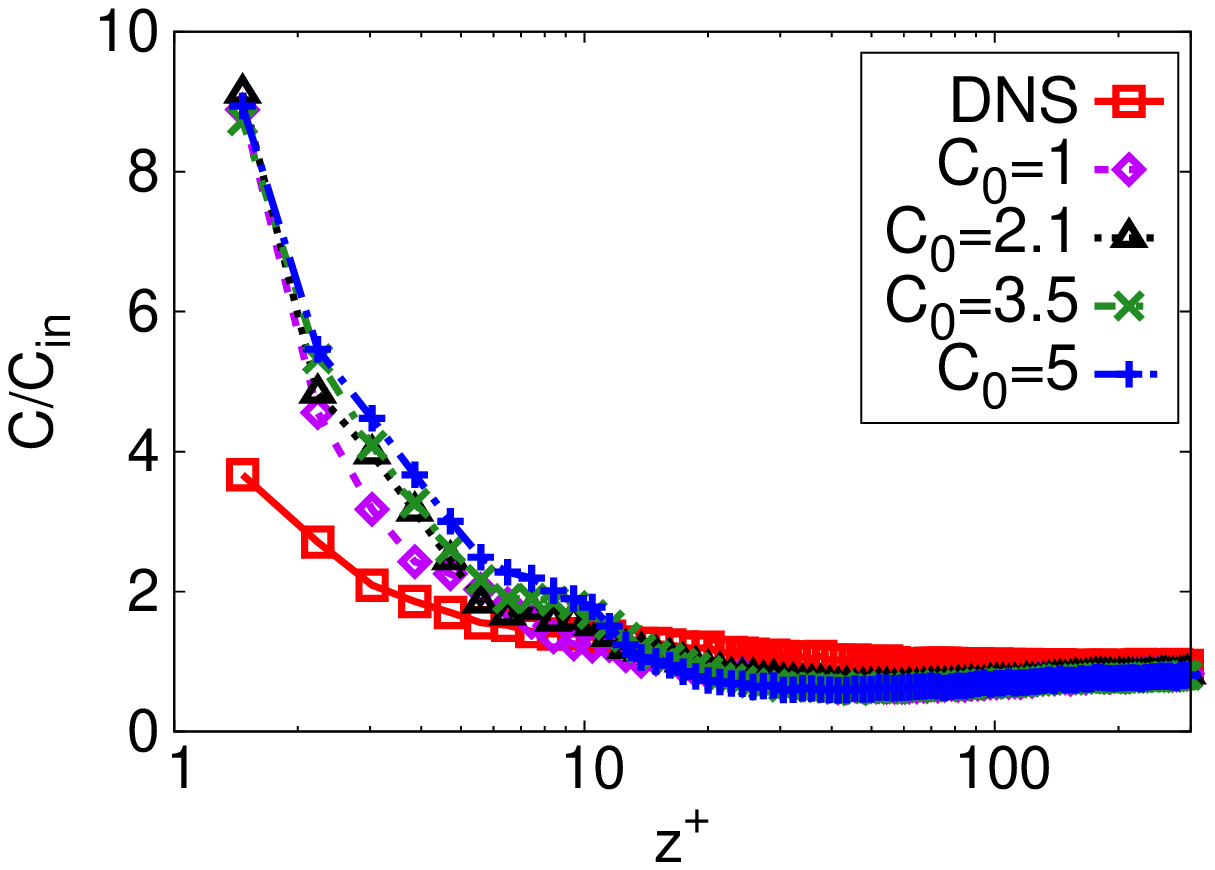}}
{\includegraphics[width=.32\textwidth]{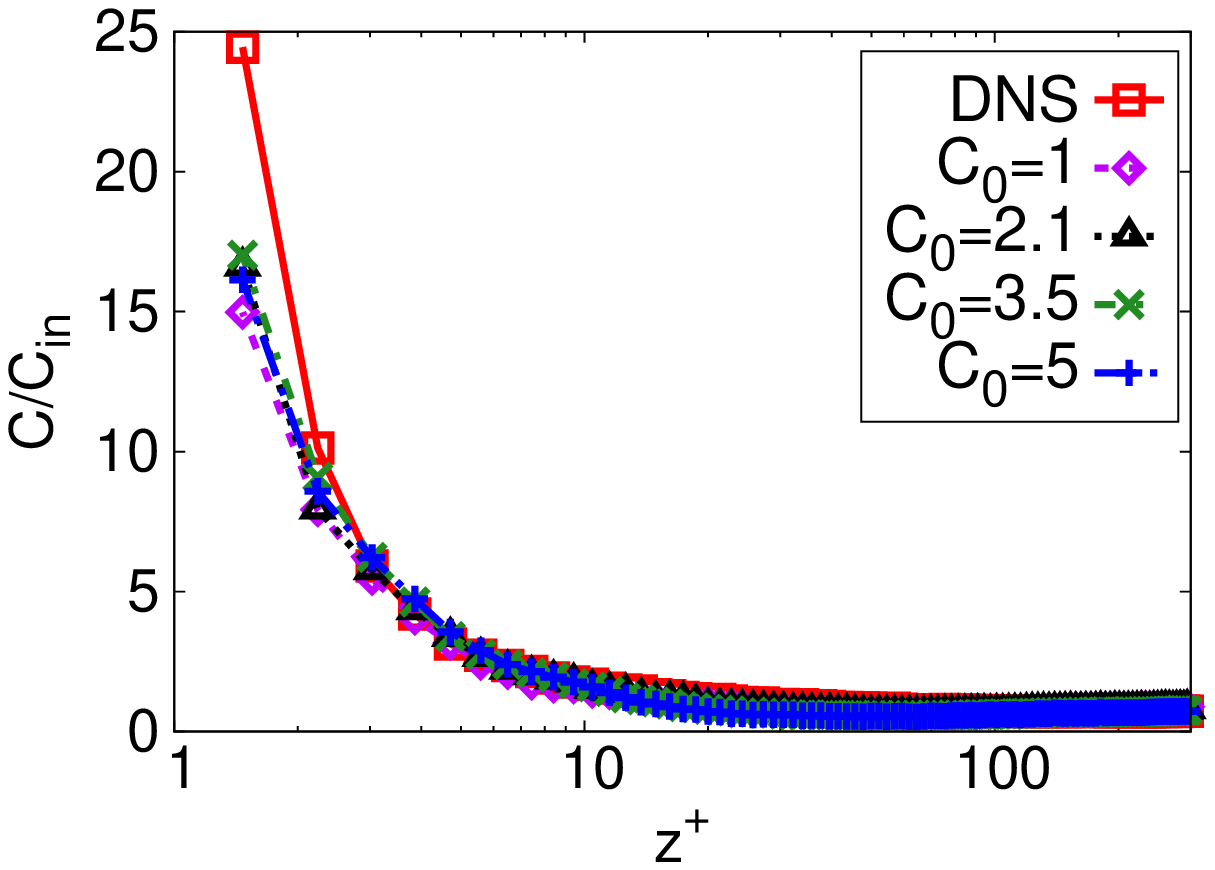}}
{\includegraphics[width=.32\textwidth]{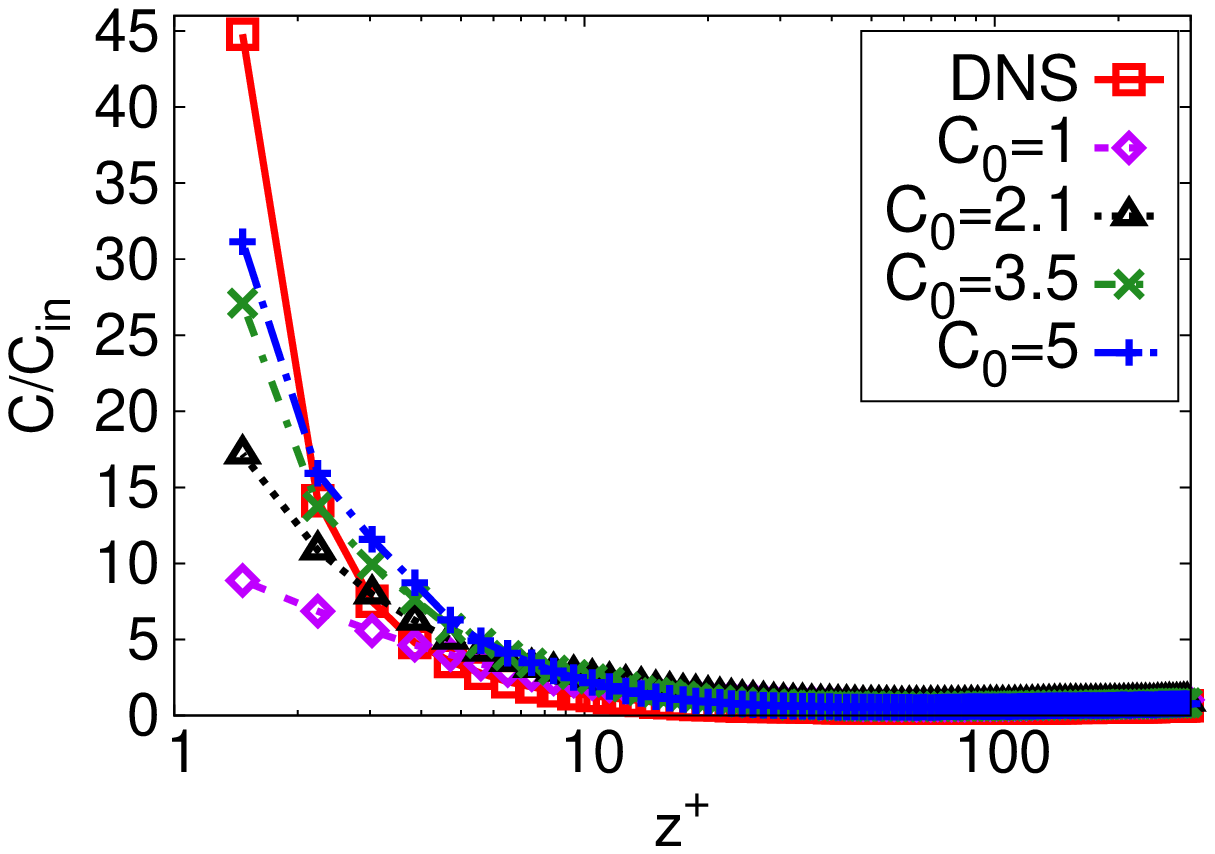}}
\caption{Effect of parameter $C_0$ on particle number density along the wall-normal coordinate
(\emph{a-priori} estimate).
Red symbols ({\color{red}{$\square$}}) refer to the DNS result, all other symbols
refer to LES results obtained with the LFMDF model.
Panels: (a) $St=1$, (b) $St=5$, (c) $St=25$.
Profiles are computed at $t^+ =2130$ after particle injection. }
\vspace{-6.9cm}
\hspace{-1.5cm} (a) \hspace{4.7cm} (b) \hspace{4.7cm} (c)
\vspace{7.0cm}
\label{Fig:concentrations-C0}
\end{figure}

\begin{figure}[b]
{\includegraphics[width=.32\textwidth]{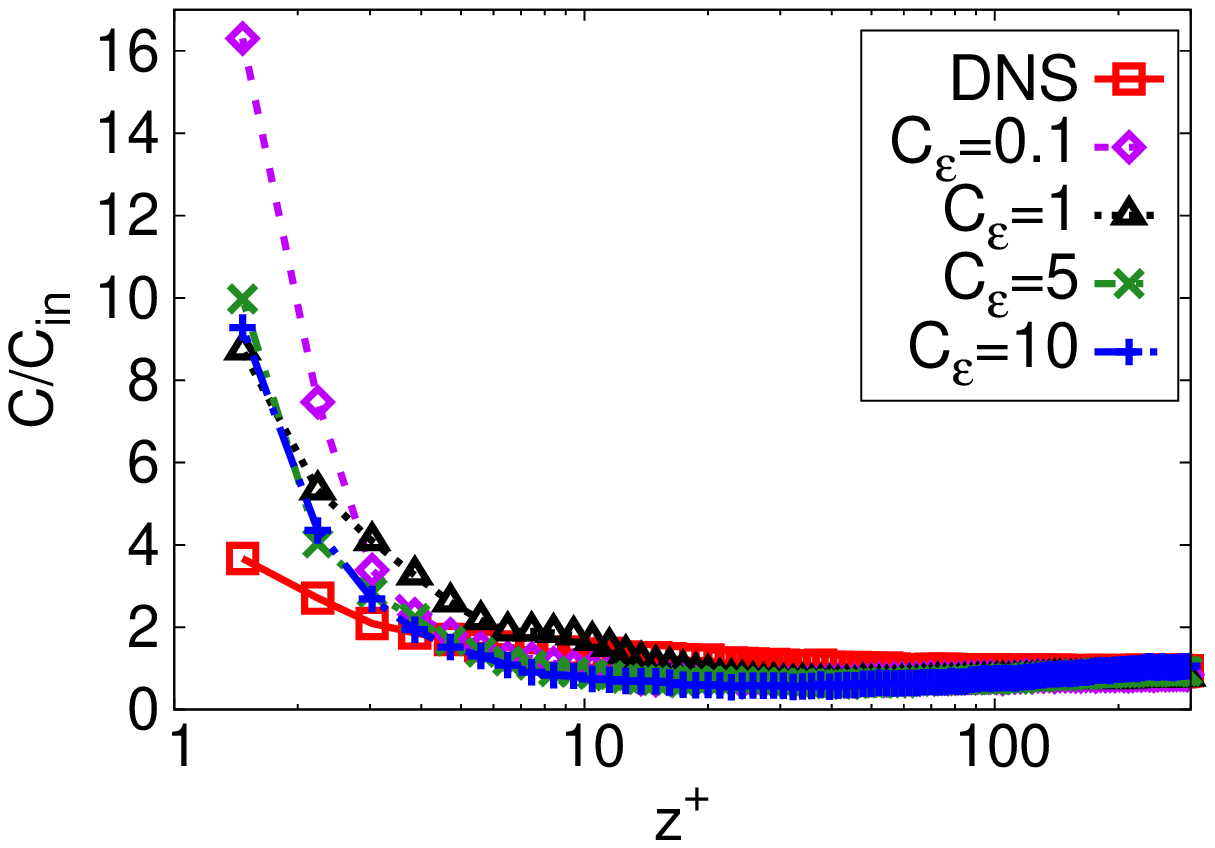}}
{\includegraphics[width=.32\textwidth]{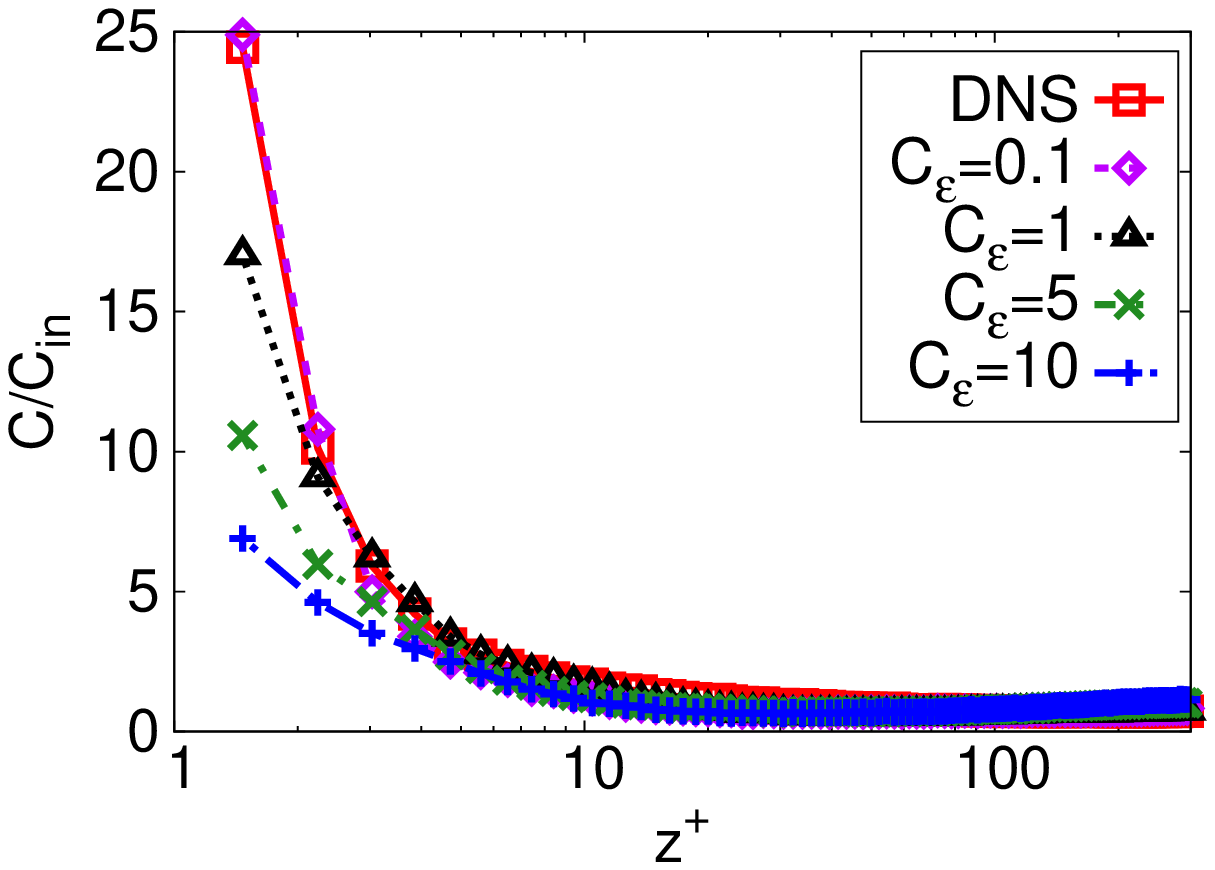}}
{\includegraphics[width=.32\textwidth]{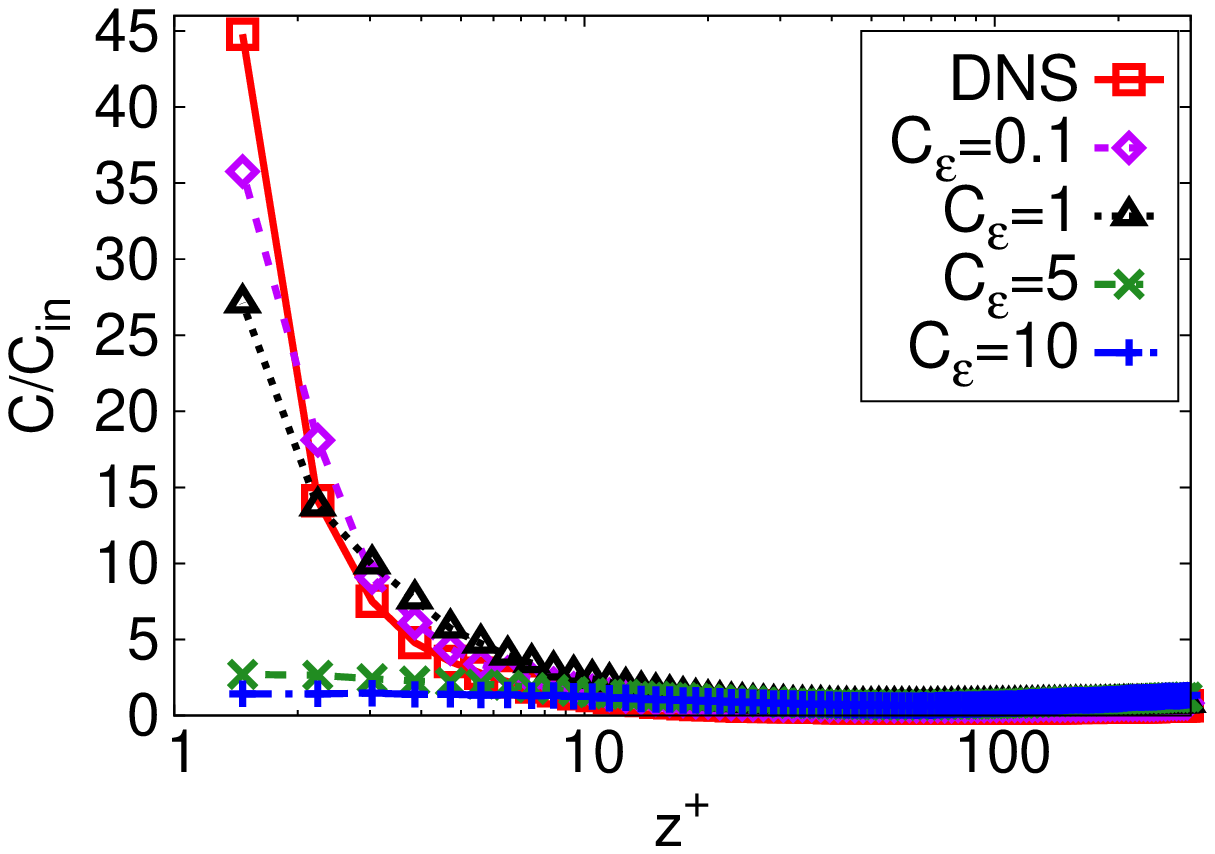}}
\caption{Effect of parameter $C_{\epsilon}$ on particle number density along the wall-normal coordinate
(\emph{a-priori} estimate).
Red symbols ({\color{red}{$\square$}}) refer to the DNS result, all other symbols
refer to LES results obtained with the LFMDF model.
Panels: (a) $St=1$, (b) $St=5$s, (c) $St=25$.
Profiles are computed at $t^+ =2130$ after particle injection.}
\vspace{-6.9cm}
\hspace{-1.5cm} (a) \hspace{4.7cm} (b) \hspace{4.7cm} (c)
\vspace{7.0cm}
\label{Fig:concentrations-CE}
\end{figure}

Figure \ref{Fig:concentration-nok} shows the particle number density profiles along the
wall-normal coordinate for different Stokes numbers.
Two different formulations of the proposed LFMDF model are tested:
The simplified \textit{LFMDF1} formulation, and the complete \textit{LFMDF2} formulation
(see Sec. \ref{sec:equiv_stoc_syst}).
In both formulations we use $C_0=2.1~,~C_{\epsilon}=1$, $\beta = 0.8$.
The \textit{LFMDF1} predictions (dark magenta profiles) deviate substantially from
the reference DNS results (red profiles)
for all Stokes numbers: This is due, of course, to the assumption of isotropic velocity
fluctuations on which the LFMDF1 formulation is based. 
On the other hand, the \textit{LFMDF2} formulation, which has a
more complete diffusion term, leads to improved predictions (black profiles), especially for the two
larger Stokes numbers: $St=5$, panel (b); and $St=25$, panel (c).
Discrepancies, however, are still evident and lead to a significant over-estimation
(under-estimation) of particle accumulation in the viscous sub-layer for the smaller $St=1$
(large $St=25$) particles, as shown in Fig. \ref{Fig:concentration-nok}(a) and in
Fig. \ref{Fig:concentration-nok}(c) respectively.
The main reason is that the closure
of the LMFDF2 formulation involves
two parameters, $C_0$ and $C_\epsilon$, which are known to be quite
sensitive to the characteristic features of both the turbulent
flow and the numerical approach.
For instance, turbulent theory leads to set $C_0=2.1$ for stochastic models in homogeneous
flows \cite{pope2000turbulent}, whereas numerical simulations of wall-bounded flows in the RANS framework
suggest to set $C_0=3.5$ \cite{Min_99}.
In this study, we exploit DNS to obtain \emph{a priori} estimates of the two model constants. 
We remark that our purpose is not to find optimal values for $C_0$ and $C_\epsilon$, but rather to
quantify the sensitivity of the model to a change in the value of these constants.
Figure \ref{Fig:concentrations-C0} shows the number density profiles obtained at varying $C_0$
(while keeping $C_\epsilon$ constant and equal to 1).
This figure shows that $C_0$ has a significant influence on particle wall-normal accumulation only
for large-inertia particles (high Stokes numbers), and suggests that $C_0=3.5$ provides the best
predictions over the range of Stokes numbers considered here. 
We performed a similar analysis to estimate  $C_{\epsilon}$ while keeping $C_0$ constant (and equal to 3.5).
Results are shown in Fig. \ref{Fig:concentrations-CE} and demonstrate that $C_{\epsilon}$ affects
particle spatial distribution at all Stokes numbers. In particular, we observe higher accumulation of particles
at the wall for smaller values of $C_{\epsilon}$. This finding indicates that the diffusion term is at least
as important as the drift term in the present flow configuration.
Based on this comparison, we select $C_{\epsilon}=0.1$ to calibrate the LFMDF model.

A combined analysis of Figs. \ref{Fig:concentrations-C0} and \ref{Fig:concentrations-CE} indicates
that, regardless of the value considered for $C_0$ and $C_{\epsilon}$, the near-wall concentration
of small inertia particles (represented by the $St=1$ particles in this study) is always overestimated
by the LFMDF2 model, whereas the opposite occurred with the LFMDF1 model (see Fig.
\ref{Fig:concentration-nok}a). For such particles, therefore, the critical modelling issue in order to
retrieve the correct physical behaviour seems to be the closure of the diffusion term.
We remark here that particles with small inertia are subject to a weaker turbophoretic wallward drift
and tend to remain more homogeneously distributed within the flow domain
\cite{Sol_09,Mar_02}. As a consequence, the
instantaneous Eulerian statistics that can be extracted from local particle ensemble averages may
exhibit significant statistical errors in the near-wall region, where the control volumes to which
averaging is applied become smaller and smaller. This source of error becomes less important
as particle inertia increases, namely as particle accumulation in the near-wall region increases
with $St$.

The key quantity for a correct evaluation of the diffusion term
is the kinetic energy ratio $\widehat{k}_{SGS} / k_{SGS}$. If $\widehat{k}_{SGS}$ is computed from Eq. (\ref{eq:kappa}),
which implies Lagrangian ensemble averaging, then it will be affected by the resulting statistical error.
\begin{figure}[b]
{\includegraphics[width=.32\textwidth]{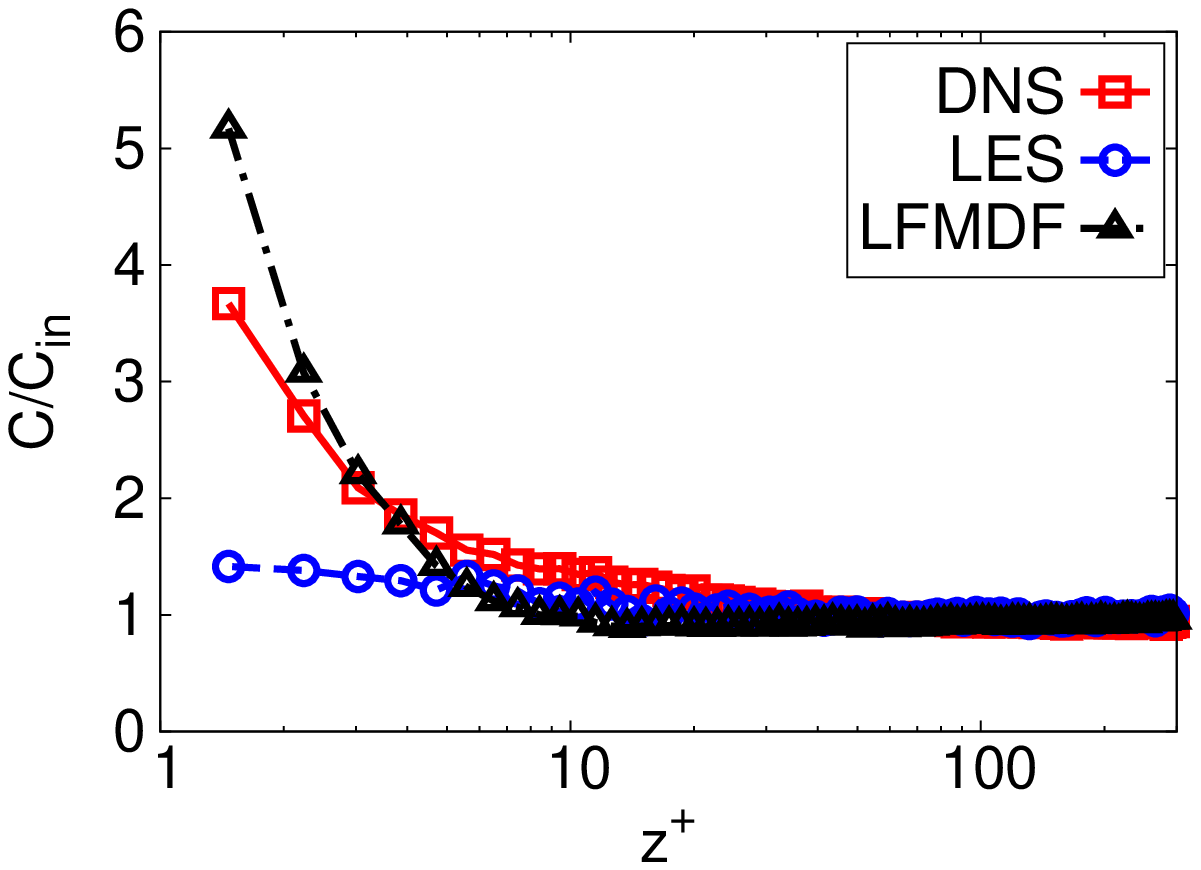}}
{\includegraphics[width=.32\textwidth]{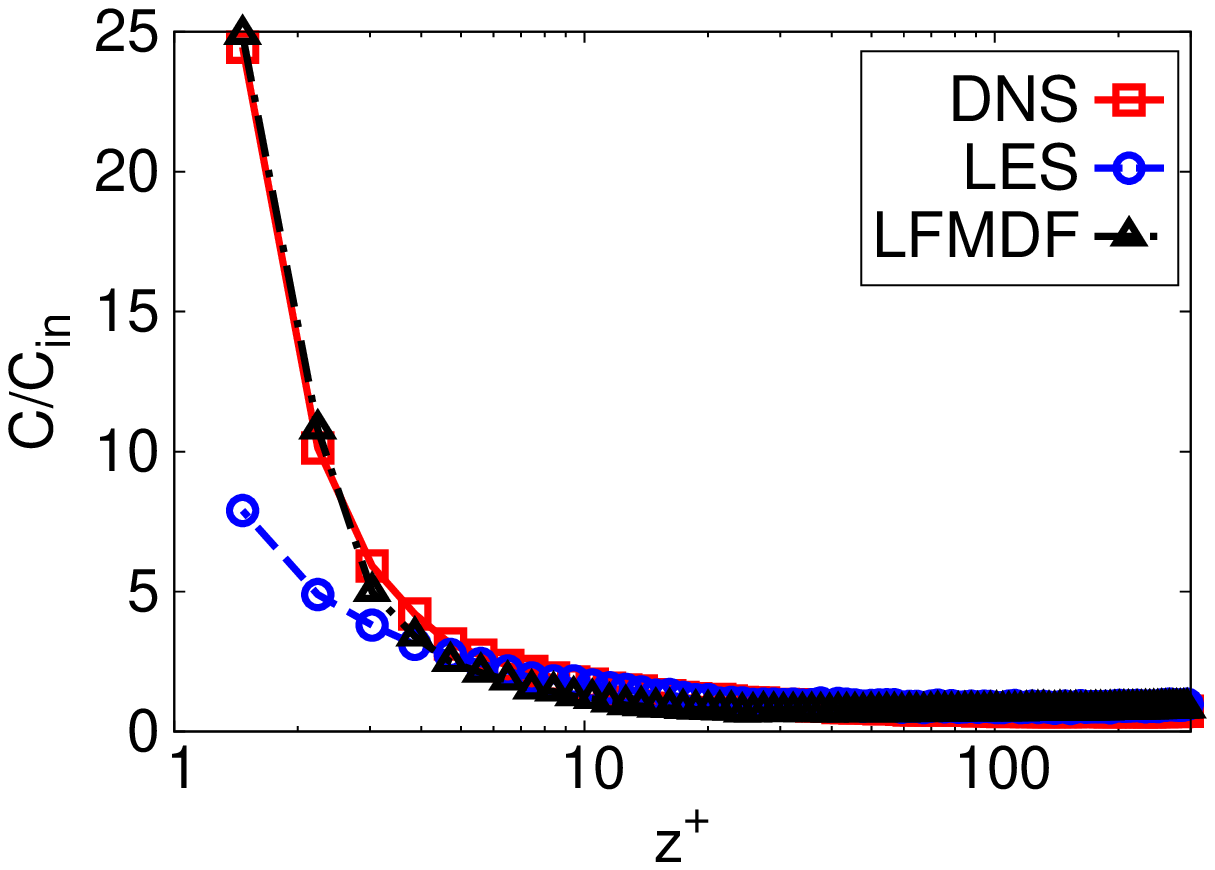}}
{\includegraphics[width=.32\textwidth]{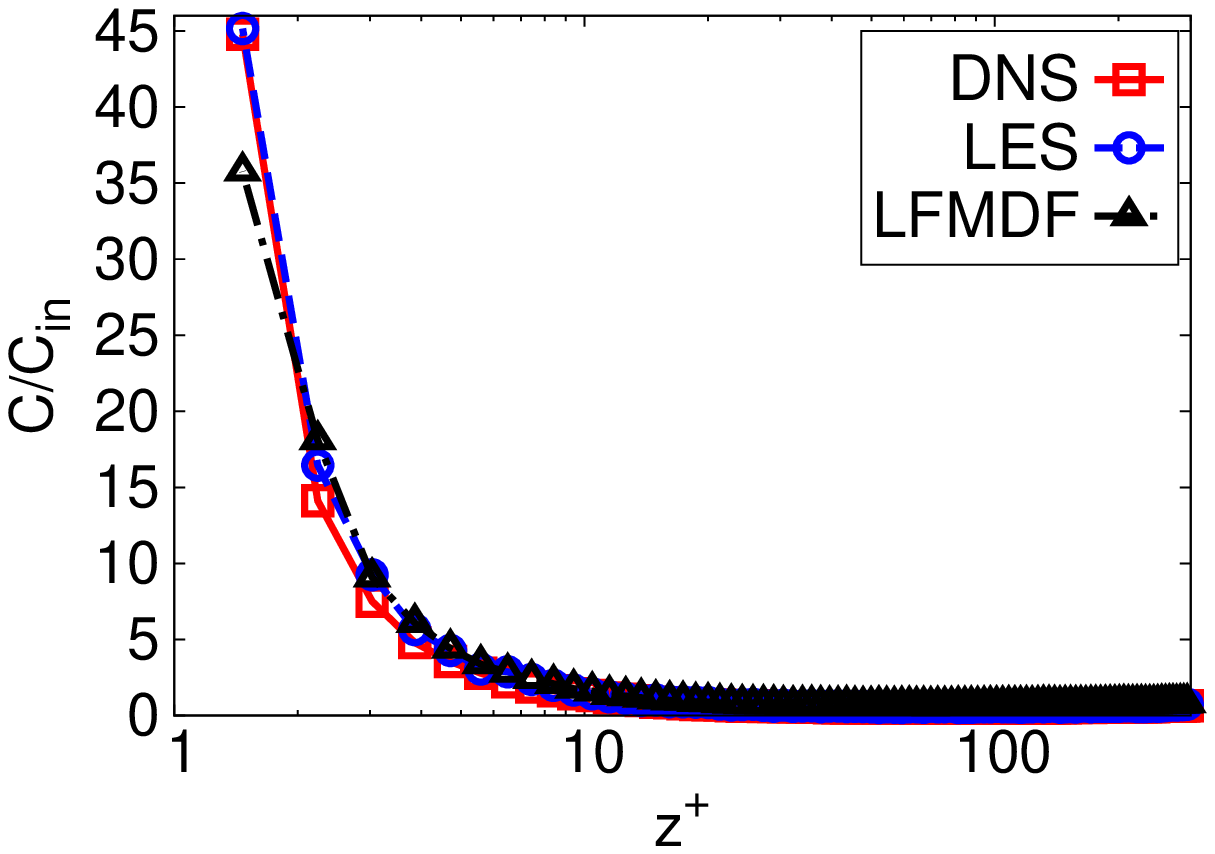}}
\caption{Comparative assessment of the LFMDF model with Eq. (\ref{eq:st1}): Predictions of
the instantaneous particle number density at varying Stokes numbers
({\color{black}{$\triangle$}})  are compared with
DNS results ({\color{red}{$\square$}}) and with LES results with no particle SGS
model ({\color{blue}{$\circ$}}).
Panels: (a) $St=1$, (b) $St=5$, (c) $St=25$.
Profiles are computed at $t^+ =2130$ after particle injection. }
\vspace{-6.9cm}
\hspace{-1.5cm} (a) \hspace{4.7cm} (b) \hspace{4.7cm} (c)
\vspace{7.0cm}
\label{Fig:concentrations}
\end{figure}
To improve the model, we propose a new formulation to evaluate $k_{SGS}$, which is slightly different from Eq. (\ref{eq:def-eps}):
\begin{equation}
k_{SGS}=
\tau(U_{s,i},U_{s,i})=\frac{1}{2}
\sum_{i=1}^{3}
\left[ \widetilde{U_{s,i}^2}-(\widetilde{U_{s,i}})^2 \right]~.
\label{eq:st1}
\end{equation}
In the limit of $N_{pc} \to \infty$, Eq. (\ref{eq:st1}) is equivalent to Eq. (\ref{eq:kappa}), but is
expected to decrease the variance of the model estimations for finite values of $N_{pc}$ at
small Stokes numbers. In the following, results for the $St=1$ particles refer to calculations
performed using this new formulation, unless otherwise stated.
In particular, Fig. \ref{Fig:concentrations} shows the comparison of the LFMDF results for particle
number density.
For completeness, also the LES results
without particle SGS model are included.
The overshoot of particle accumulation at the wall for $St=1$ is strongly reduced with respect to
the predictions reported in Figs. \ref{Fig:concentrations-C0} and \ref{Fig:concentrations-CE}, and there is a nearly perfect
match with the DNS profile for the intermediate-inertia particles ($St=5$, Fig. \ref{Fig:concentrations}b).
As expected, wall accumulation at large Stokes numbers is unaffected.
We remark that the values of particle number density within a distance of few wall units from the wall
are very noisy even in DNS \cite{prevel2013direct}: This implies that the only relevant information one
can extract from the viscous sublayer portion of the profiles shown in Fig. \ref{Fig:concentrations} is
just the trend in model performance at varying particle inertia.

\begin{figure}[t]
{\includegraphics[width=0.48\textwidth]{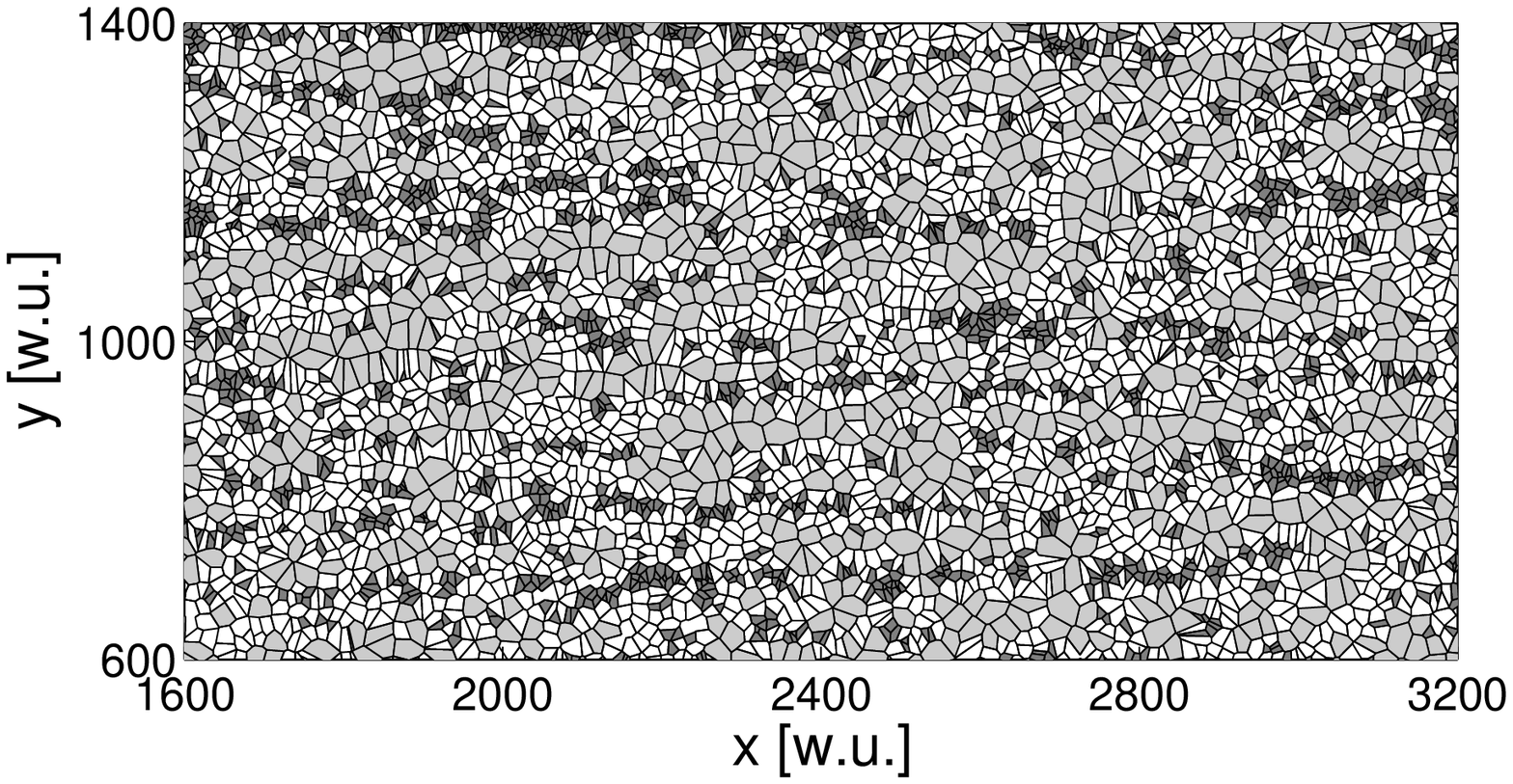}}\\
{\includegraphics[width=0.48\textwidth]{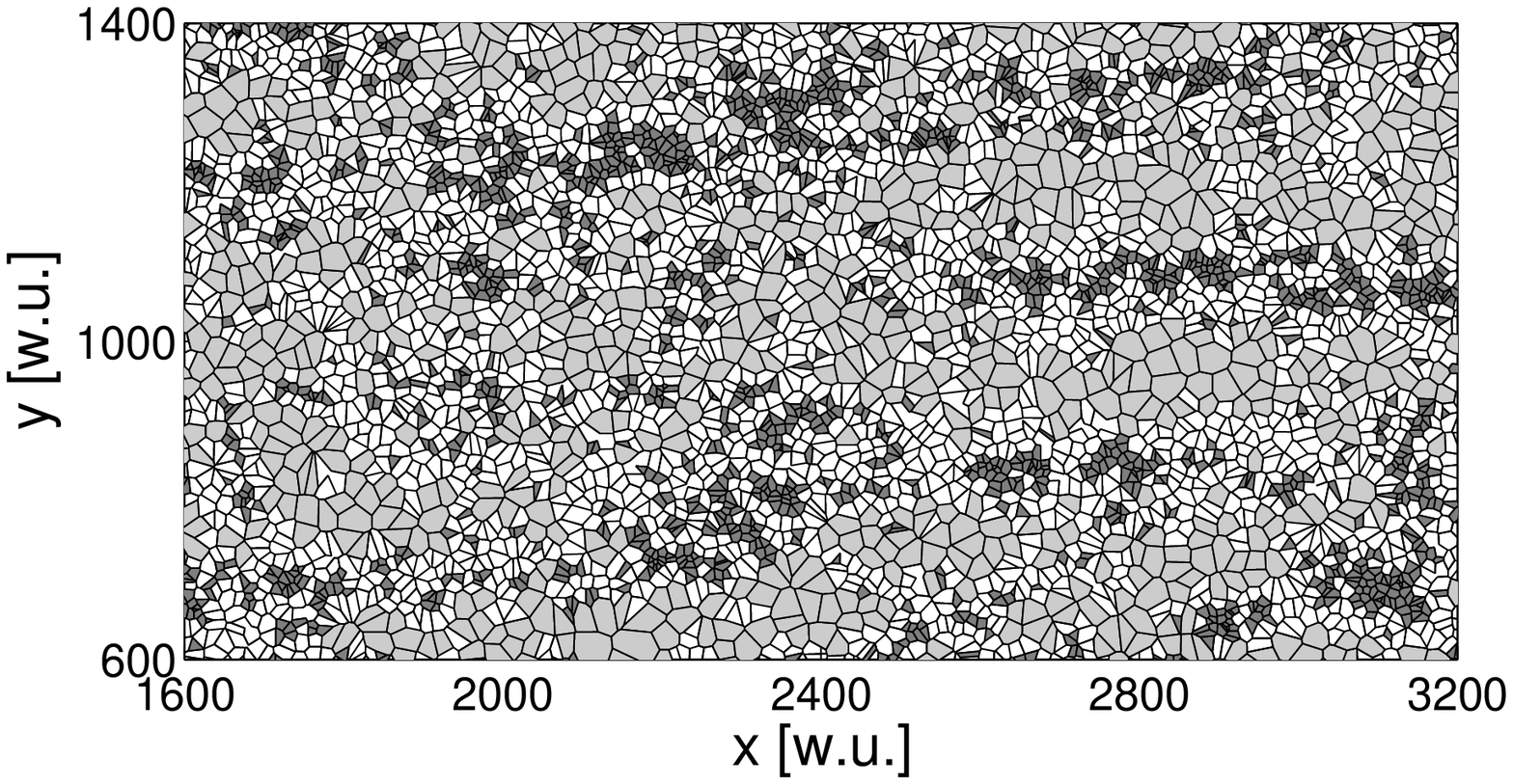}}\\
{\includegraphics[width=0.48\textwidth]{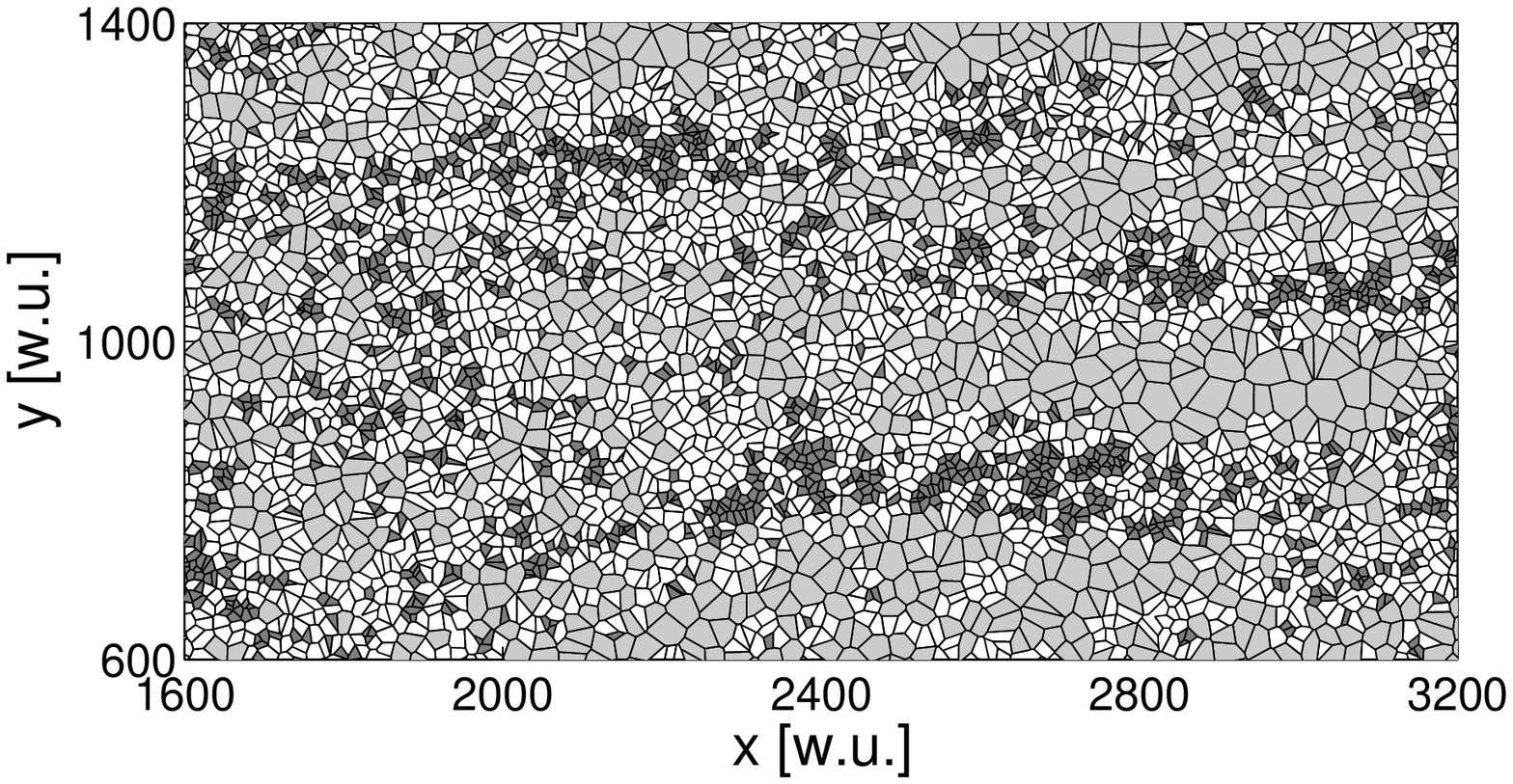}}
\caption{Vorono\"i tessellation for the $St=5$ particles on a wall-parallel fluid slab ($1 \leq z^+ \leq 5$)
at time $t^+ =2130$ after particle injection.
Particle clusters are in dark gray, voids are in ligth gray.
Panels: (a) DNS, (b) LES with no particle SGS model, (c) LES with the calibrated LMFDF model. 
}
\vspace{-15.9cm}
\hspace{-8.5cm} (a)

\vspace{3.4cm}
\hspace{-8.5cm} (b)

\vspace{3.4cm} 
\hspace{-8.5cm} (c)

\vspace{6.5cm}
\label{Fig:voronoi1}
\end{figure}

\begin{figure}[]
{\includegraphics[width=0.47\textwidth]{figure12a.eps}}
{\includegraphics[width=0.49\textwidth]{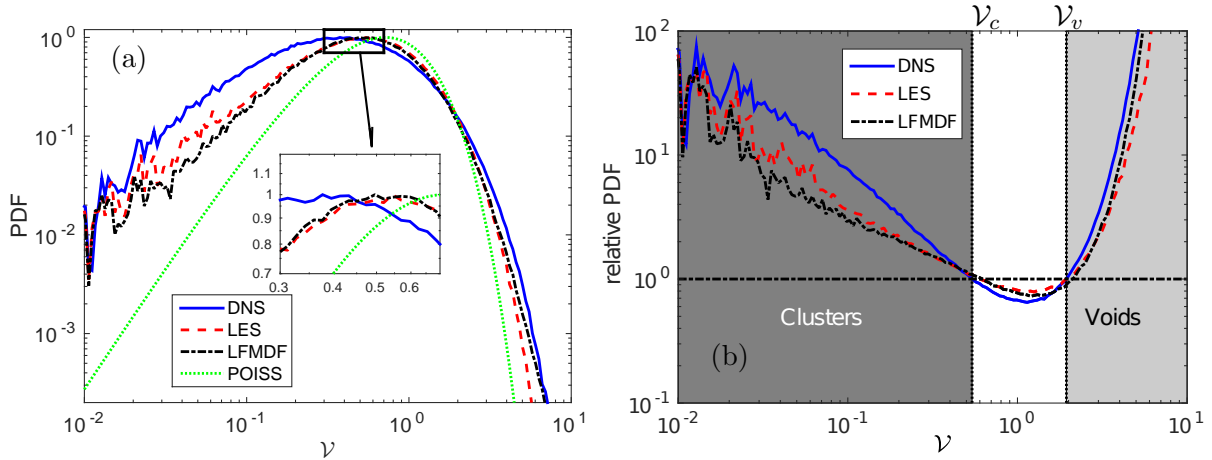}}
\caption{PDF of normalized Vorono\"i areas ($\mathcal{V} = A / \bar{A}$) of $St = 5$ particles on a
wall-parallel fluid slab located at distance $1 \leq z^+ \leq 5$ from the wall.}
\vspace{-8.3cm}
\hspace{11.1cm} $\mathcal{V}_c$ \hspace{0.45cm} $\mathcal{V}_v$

\vspace{-0.2cm} 
\hspace{-12.8cm} (a)

\vspace{3.2cm}
 \hspace{3.3cm} (b)
 \vspace{3.0cm} 
\label{Fig:voronoi2}
\end{figure}

To provide a phenomenological perspective to our discussion, we complement the statistical
description of particle wall-normal distribution with the analysis of particle clustering in the
near-wall region. As demonstrated in previous studies (see \cite{Mar_02,Pic_05,Sol_09} and references therein,
for a review), the tendency that inertial particles have to form clusters is crucial to develop
peaks of particle concentration within the flow. Therefore, a reliable particle SGS model should be
able to capture (in a statistical sense) also these phenomena.
To perform this analysis, we quantify particle clusters by means of Vorono\"i diagrams, which represent
an efficient and robust tool to diagnose and quantify clustering~\cite{Monchaux10}. 
One Voronoi cell is defined as the ensemble of points that are closer to a given particle than to
any other particle in the flow: The area of a Vorono\"i cell is therefore the inverse of the local particle
number density. In addition Vorono\"i areas are naturally evaluated around each particle and, differently
from standard box counting methods, provide a direct measure of particle preferential concentration
at inter-particle length scale~\cite{Monchaux10}.
An example of Vorono\"i diagram for the present channel flow configuration is shown in Fig. \ref{Fig:voronoi1},
which focuses on the instantaneous distribution of the $St=5$ particles within a wall-parallel fluid slab 
of thickness $1 \leq z^+ \leq 5$.  Only a portion of the $x-y$ plane is shown to highlight the presence
of the well-know particle streaks. Compared to the visualisation provided by DNS (Fig. \ref{Fig:voronoi1}a),
both LES results (with no particle SGS model in Fig. \ref{Fig:voronoi1}(b); with the LMFDF model
in Fig. \ref{Fig:voronoi1}(c), respectively) show broader particle streaks and wider inter-cluster spacing.
Clusters and voids are identified by comparing the PDF of Vorono\"i areas obtained from the simulations
to that of a synthetic random Poisson process, whose shape is well approximated
by a Gamma distribution~\cite{Monchaux10}. This comparison is shown in Fig. \ref{Fig:voronoi2}, where the Vorono\"i areas
are normalized using the average Vorono\"i area, ${\bar{A}}$ (equivalent to the inverse of the mean
particle number density), independent of the spatial organization of the particles.\\
As found previously~\cite{Monchaux10}, in the case of heavy particles,
the PDFs clearly depart from the Poisson distribution,
with higher probability of finding depleted
regions (large Vorono\"i areas) and concentrated regions
(small Vorono\"i areas), a typical signature
of preferential concentration.
In the present study, the inclusion of the LMFDF model into the LES has little effect on the prediction
of concentrated regions, and the first cross-over point, $\mathcal{V}_c$,
representing the threshold value below which Vorono\"i areas are considered to belong to
a cluster,
occurs at slightly larger values than in DNS. The model improves prediction
of depleted regions even if the second cross-over point, $\mathcal{V}_v$,
representing the threshold value above which Vorono\"i areas are considered to belong to
a void, is always well predicted.

\begin{figure}[]
{\includegraphics[width=.32\textwidth]{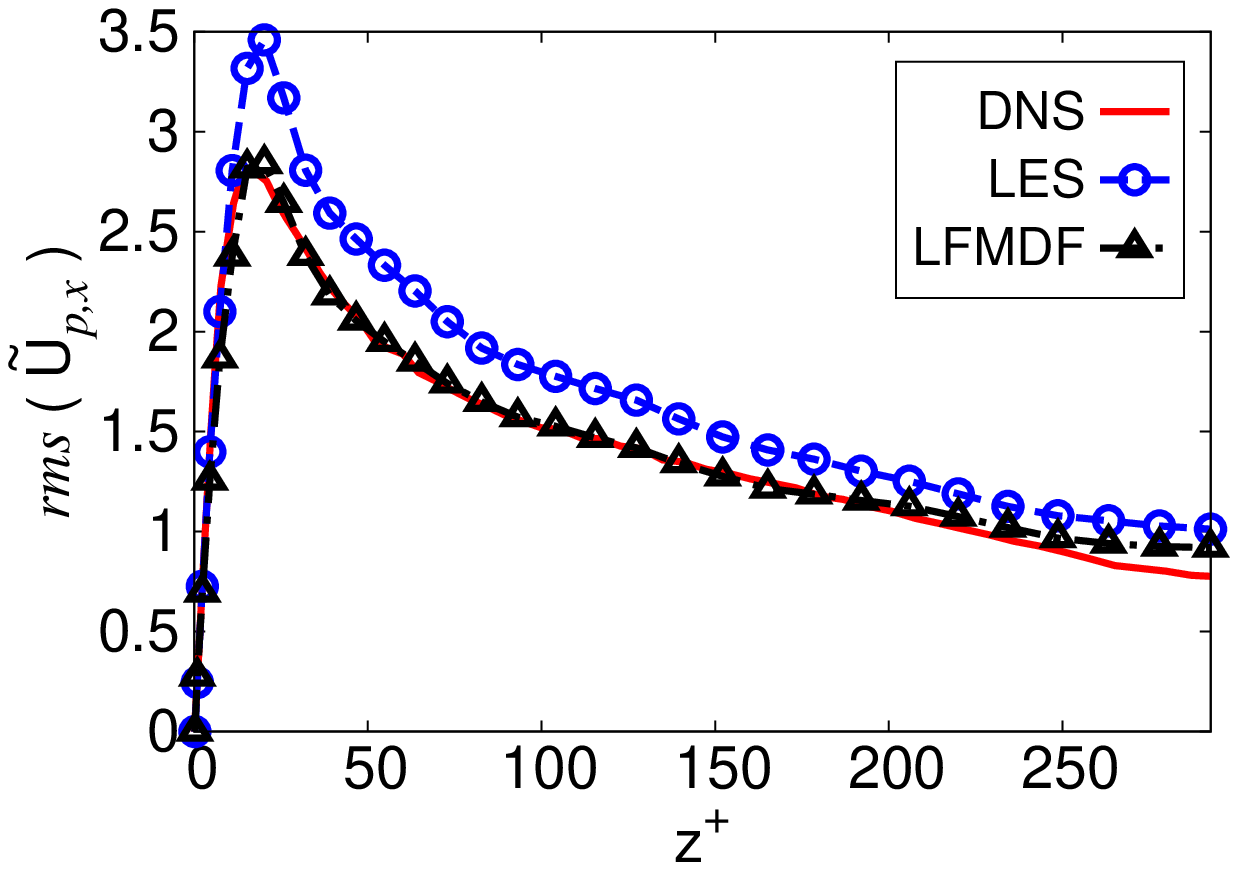}}
{\includegraphics[width=.32\textwidth]{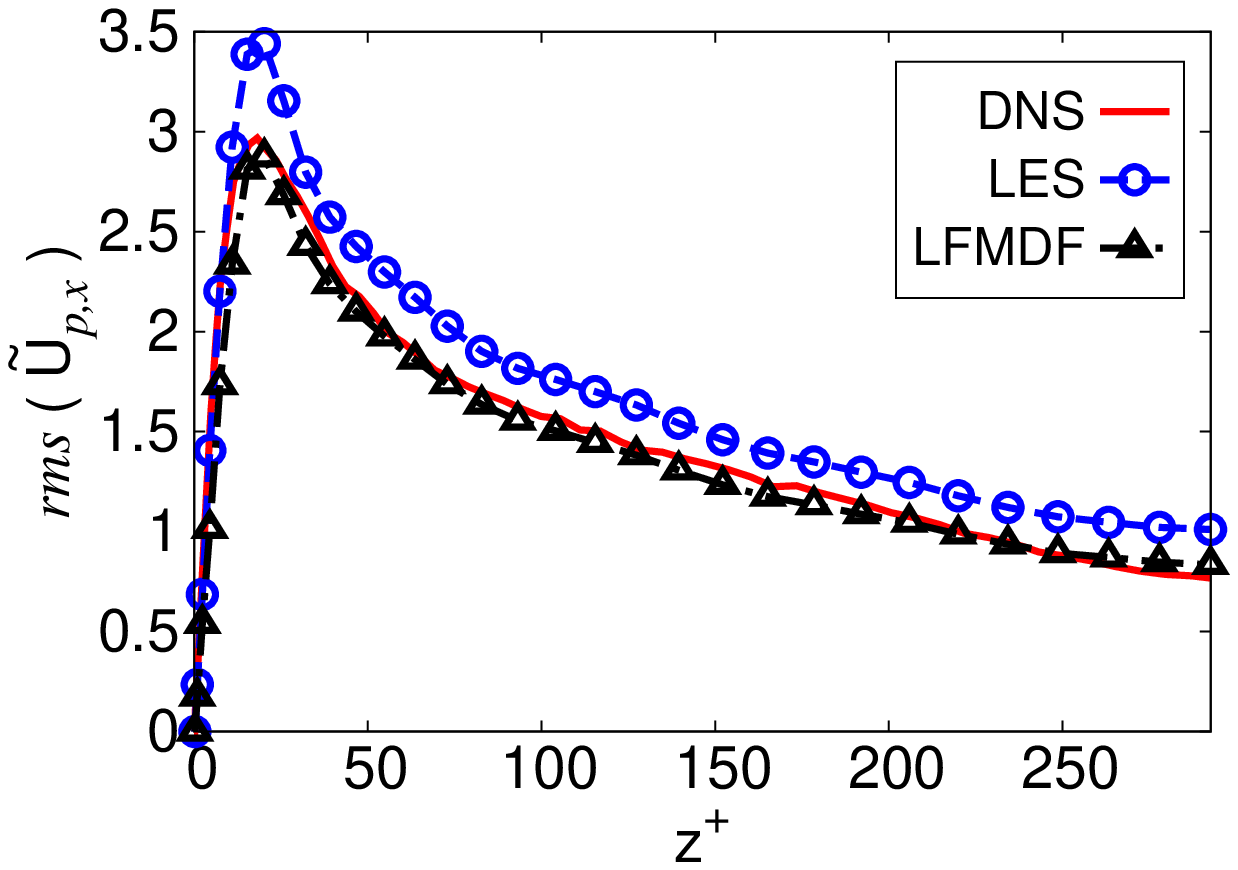}}
{\includegraphics[width=.32\textwidth]{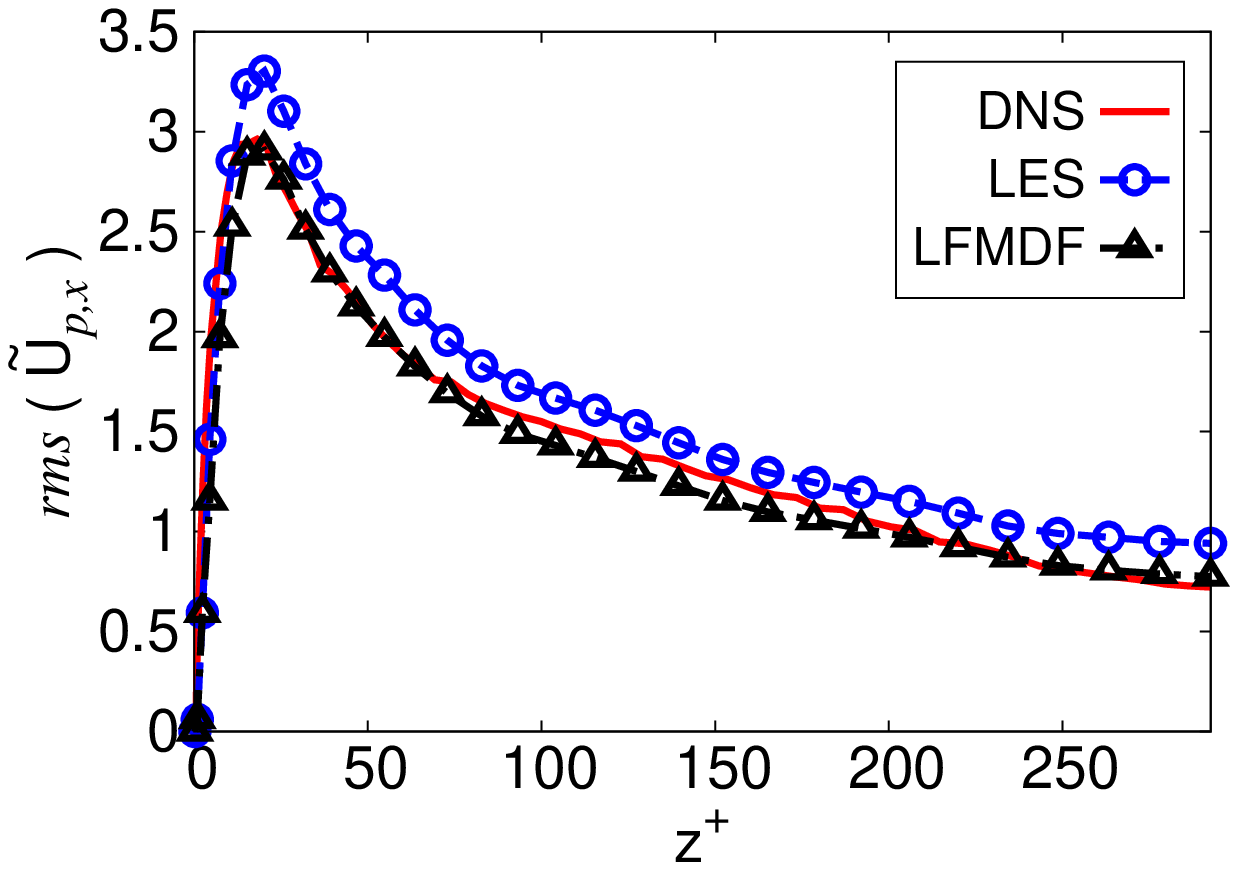}}
\vspace{10mm}
{\includegraphics[width=.32\textwidth]{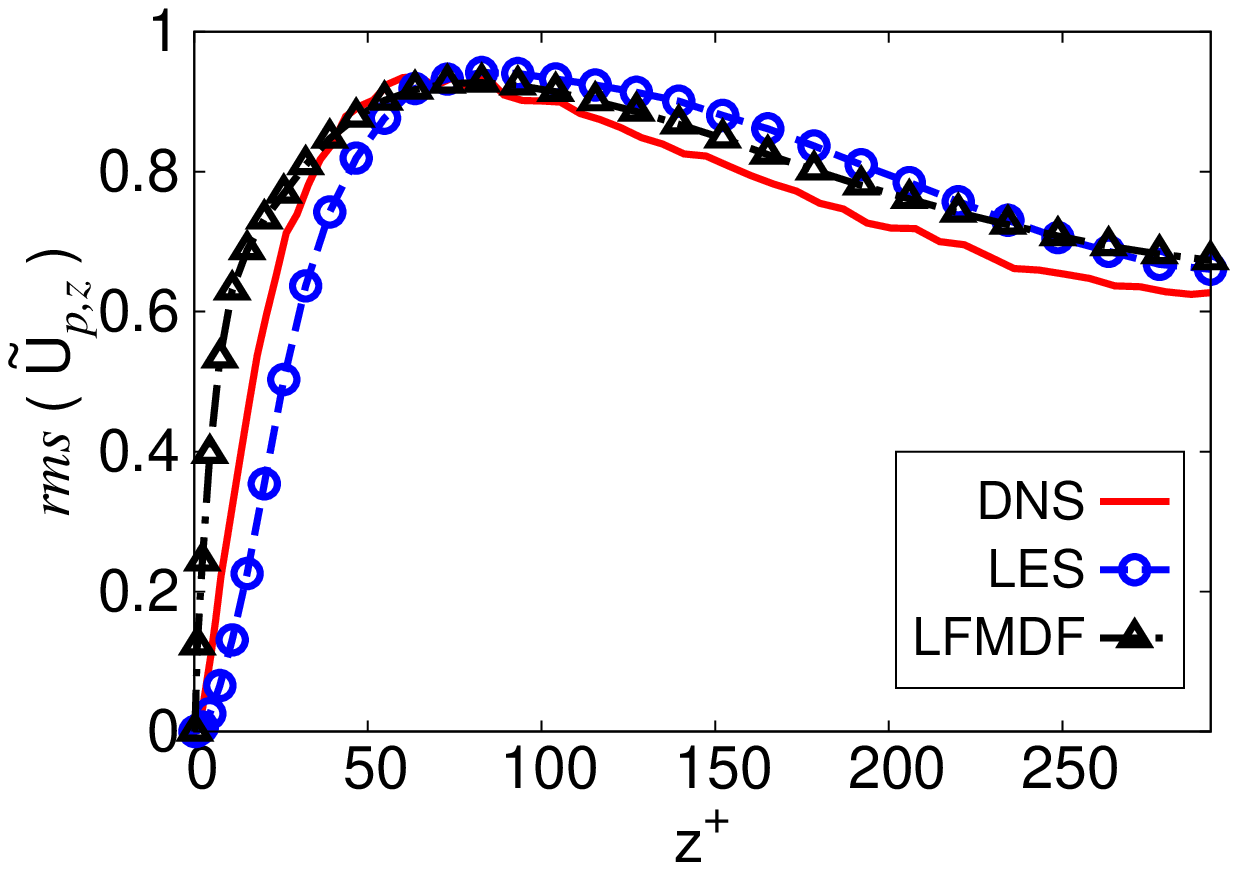}}
{\includegraphics[width=.32\textwidth]{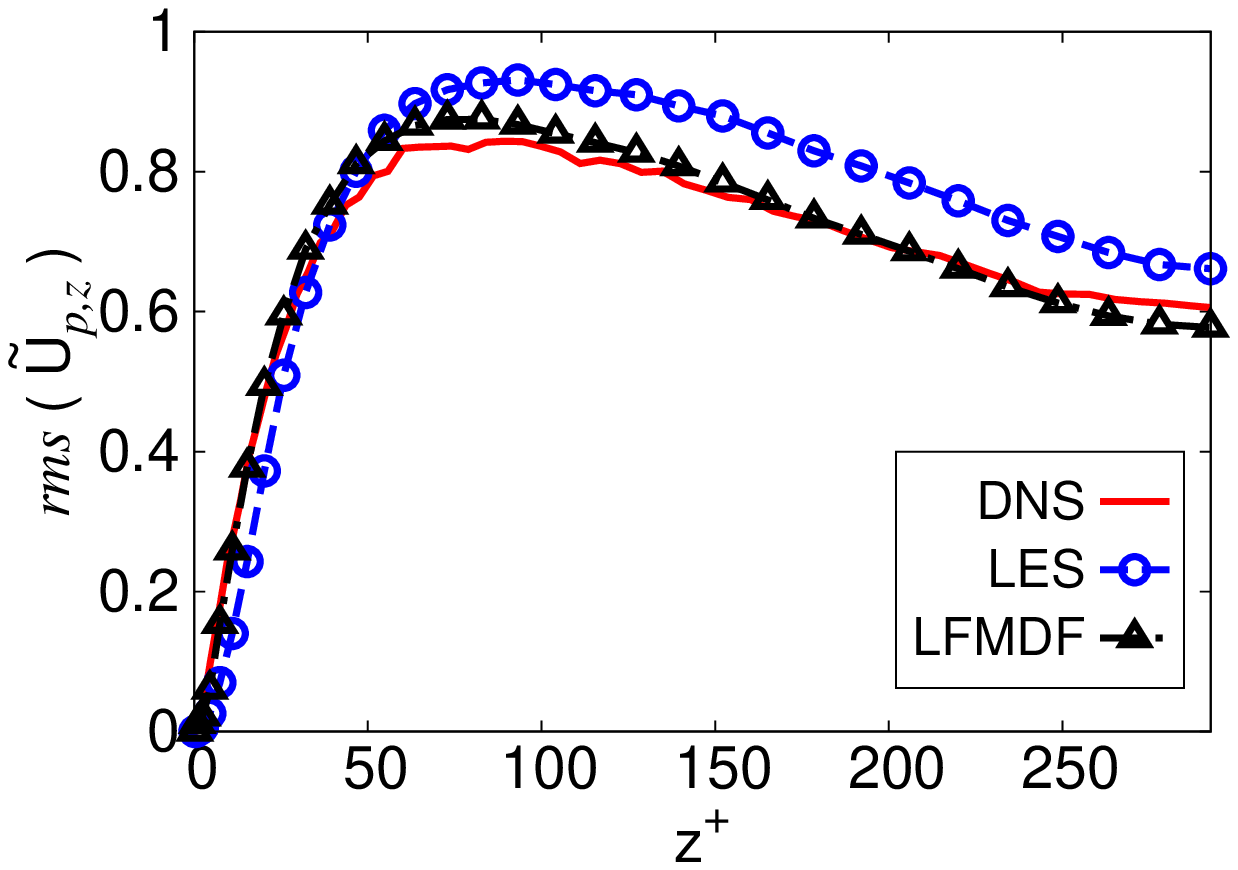}}
{\includegraphics[width=.32\textwidth]{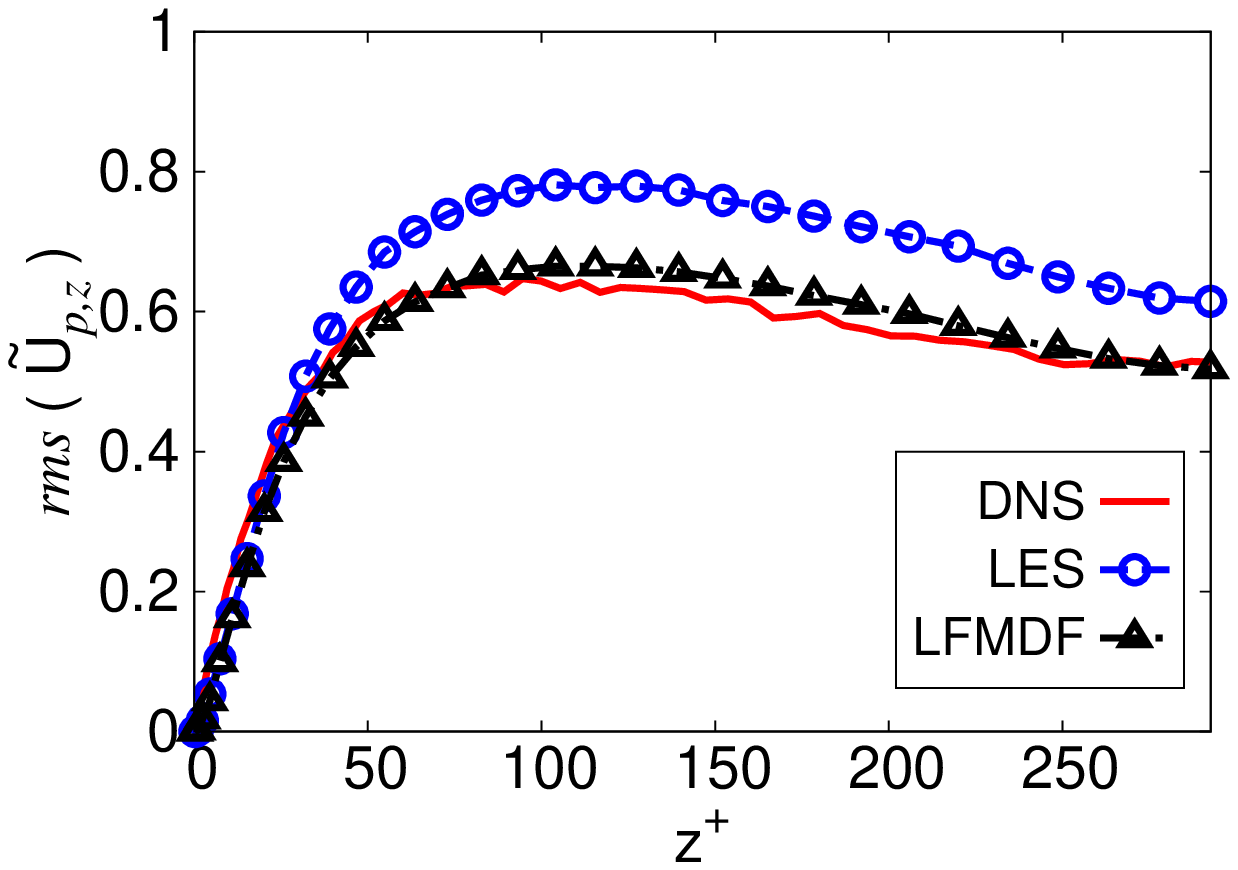}}
\vspace{-0.8cm}
\caption{Comparative assessment of the LFMDF model: Prediction of the particle velocity rms
at varying Stokes number ({\color{black}{$\triangle$}})  are compared with
DNS results ({\color{red}{------}}) and with LES results with no particle SGS
model ({\color{blue}{$\circ$}}).
Panels: (a),(d) $St=1$, (b),(e) $St=5$, (c)-(f) $St=25$; (a)-(c) streamwise component, (d)-(f) wall-normal
component. Statistics are obtained averaging over a time window $\Delta t^+ =1800$. }
\vspace{-10.3cm}
\hspace{-2.3cm} (a) \hspace{4.7cm} (b) \hspace{4.7cm} (c)

\vspace{3cm}
\hspace{-2.3cm} (d) \hspace{4.7cm} (e) \hspace{4.7cm} (f)
\vspace{6.5cm}
\label{Fig:rms-part}
\end{figure}

To complete the LFMDF model assessment, in Fig. \ref{Fig:rms-part} we show the statistics of the
root mean square of particle velocity. In particular, we focus on the streamwise and wall-normal
components, which are the most interesting as far as particle wall transport is concerned.
It can be seen that the calibrated LFMDF improves the LES prediction for all Stokes numbers, with
just small (yet persistent) discrepancies for the wall-normal rms of the $St=1$ particles (Fig. \ref{Fig:rms-part}d).
This explains the peak of concentration observed for these particles in the number density statistics.

\section{Discussion and conclusions}

In this work, we have presented a new FDF approach to the simulation
of turbulent dispersed flows. The approach is derived from RANS- and LES-based models
that have been successfully applied to the simulation of reactive and polydispersed flows \cite{pope2000turbulent,fox2012large,Pei_06}. 
We have put forward a Lagrangian Filtered Mass Density Function (LFMDF) model that
provides the Lagrangian probability density function of the SGS particle variables and of the
fluid velocity seen by the particles.
Important features of the proposed method are that (1) at variance with reactive flows, the approach
is Lagrangian and (2) a mass density function is considered, as done in compressible flows.
The exact transport equation for the LFMDF has been derived, and a modeled transport equation for
the filtered density function has been developed using a closure strategy inspired by PDF methods.
Specifically, two different formulations have been proposed, which differ in the treatment of the SGS
scales. It has been shown that the effects of convection and polydispersity appear in a closed form.
 
The modeled LFMDF transport equation has been solved numerically using a Lagrangian Monte Carlo
scheme and considering a set of equivalent stochastic differential equations. These equations have
been discretized with an unconditionally-stable numerical scheme based on the analytical solution that
the equations admit with constant coefficients. 
This scheme is the natural extension of the one developed in the context
of RANS simulations and is the key ingredient for the treatment of multi-scale problems.
A turbulent channel flow at shear Reynolds number $Re_\tau=300$ based on the channel half height
has been simulated and the results yield by the  LFMDF method in conjunction with LES have been
compared with those provided by large-eddy simulations in which
no SGS model is included in the particle equations.
To provide a numerical experiment as reference, results from DNS of the
same flow configuration have been considered as well. 
It is important to remark here that Reynolds number effects on the considered statistics
are expected to be marginal up to $Re_\tau \simeq 900$ \cite{geurts2012ideal}, so that present results can be considered reliable below such threshold value.

The convergence of the Monte Carlo simulations and the consistency of the LFMDF formulation in
the fluid-tracer limit have been assessed by comparing particle number density and low-order velocity
moments with those obtained from in the purely Eulerian framework. The good agreement of duplicate
(Eulerian and Lagrangian) fields demonstrates that the model can safely be applied in the case of
particles with small or negligible inertia.
We were also able to quantify the effect that the number of particles needed to compute the statistical
observables of interest (especially the number density distribution) may have.

The \emph{ a priori } assessment made against DNS allowed us to calibrate the values of the model coefficients
for the specific channel flow parameters considered in the present study. 
Even without dynamic calibration of the coefficients, 
the \emph{a posteriori } assessment made against DNS
and no-model LES show improved predictions of particle statistics (e.g. particle number density
along the wall-normal coordinate and particle velocity fluctuations), especially at intermediate Stokes
numbers.
In spite of this, however, it should be noted that the LFMDF is a purely statistical method, and therefore
can not recover much as far as turbulent coherent structures are concerned.

In our opinion, the LFMDF formulation presented in this paper provides a rigorous and physically-sound
approach to the large-eddy simulation of turbulent dispersed flows. 
While we believe it should be used as the natural framework to develop Lagrangian sub-grid models for
the dispersed phase, we are also aware that there is room for further improving the quality and predictive
capabilities of the model. A first step would be the development of a dynamic procedure to determine
the optimal values of the model coefficients, possibly as a function of the particle Stokes number.
Another improvement could be represented by the implementation of higher order closures in the
Langevin equation for the fluid velocity seen by the particles. Finally, it would be very useful to
implement low-$Re$ corrections to better capture the near-wall behaviour of the statistics: This
should improve the model predictions at relatively low particle inertia (e.g. $St=1$
in the present study).

\begin{acknowledgments}
SC warmly thanks J.-P. Minier for the contribution in the development of the formalism.
\end{acknowledgments}

\appendix

\section{Weak first-order Numerical scheme}
\label{app:scheme}
The analytical solution to the Eqns. (\ref{eq:sdeXp})-(\ref{eq:sde}) can be obtained with constant coefficients, resorting to It\^o's calculus in combination with the method of the variation of constants. Let us consider the fluid velocity seen by the particles, for instance.
One seeks a solution of the form $U_{s,i}(t)=H_i(t)\exp(-t/T_i)$, where $H_i(t)$ is a stochastic process
defined by (indicating $T_{L,i}^*$ with $T_i$ for ease of notation): \begin{equation} \label{eq:H}
dH_i(t)=\exp(t/T_i)[C_i\,dt + \Check{B}_i\,dW_i(t)],
\end{equation}
that is, by integration on a time interval $[t_0,t]$ ($\Delta t=t-t_0$),
\begin{equation} 
\begin{split}
U_{s,i}(t) = U_{s,i}(t_0)& \exp(-\Delta t/T_i)+ C_i 
             \,T_i\,[1-\exp(-\Delta t/T_i)] \\
+ & \Check{B}_i \exp(-t/T_i)
    \int_{t_0}^{t}\exp(s/T_i)\,dW_i(s), 
\end{split}
\end{equation}
where $\Check{B}_i=B_{ii}$ since $B_{ij}$ is a diagonal matrix.
The derivation of the weak first-order scheme is now rather straightforward since
the analytical solutions to Eqns. (\ref{eq:sdeXp})-(\ref{eq:sde}) with constant coefficients have been already calculated. Indeed, the Euler scheme (which is a weak scheme of order $1$) is simply obtained by freezing the coefficients at the beginning of the time interval $\Delta t = [t_n,t_{n+1}]$. Let $Z_i^n$ and $Z_i^{n+1}$ be the approximated values of $Z_i(t)$ at time $t_n$ and $t_{n+1}$, respectively. The Euler scheme is then simply
written by using the expression reported in Table \ref{tab:exa} and expressing the stochastic integrals through
the Choleski algorithm, as reported in Table \ref{tab:matcov_exa}. The second-order scheme is
based on a prediction-correction algorithm, in which the prediction step is the first-order scheme of equations (\ref{eq:first_order_scheme1})-(\ref{eq:first_order_scheme3}) and the corrector step is generated by a Taylor expansion under the assumption
that the acceleration terms vary linearly with time \cite{Pei_06}.
\begin{table}[htbp]
\caption{Analytical solutions to system (\ref{eq:sde}) for
time-independent coefficients.}
\hrule
\begin{align}
& x_{p,i}(t) = x_{p,i}(t_0)
  + U_{p,i}(t_0)\tau_p  [1-\exp(-\Delta t/\tau_p)]
  + U_{s,i}(t_0)\,\theta_i \{T_i[1-\exp(-\Delta t/T_i)] \notag \\ 
&  \hspace*{9mm} + \tau_p[\exp(-\Delta t/\tau_p)-1]\} 
  + [C_i\,T_i]
    \{\Delta t-\tau_p[1-\exp(-\Delta t/\tau_p)] 
 - \theta_i (T_i[1-\exp(-\Delta t/T_i)] \notag \\ 
&  \hspace*{9mm} +      \tau_p[\exp(-\Delta t/\tau_p)-1])\} 
     + \Omega_i(t) \label{eq:xpa_exa} \\
&  \hspace*{9mm} \text{\quad with \quad} \theta_i = T_i/(T_i-\tau_p)
  \notag \\
& U_{p,i}(t) = U_{p,i}(t_0)\exp(-\Delta t/\tau_p)
  + U_{s,i}(t_0)\,\theta_i
    [\exp(-\Delta t/T_i)-\exp(-\Delta t/\tau_p)] \notag \\ 
&  \hspace*{9mm} + [C_i\,T_i]
    \{[1-\exp(-\Delta t/\tau_p)]-\theta_i
      [\exp(-\Delta t/T_i)-\exp(-\Delta t/\tau_p)]\} 
 + \Gamma_i(t) \label{eq:Upa_exa} \\
& U_{s,i}(t) = U_{s,i}(t_0)\exp(-\Delta t/T_i)
  + C_i\,T_i[1-\exp(-\Delta t/T_i)]
  + \gamma _i(t) \label{eq:Ufa_exa} \\ \notag \\
& \text{\underline{The stochastic integrals $\gamma _i(t),\;\Gamma
_i(t),\;\Omega _i(t)$ are given by:}}\notag \\
& \quad \gamma _i(t) = \Check{B}_i\exp(-t/T_i)
  \int _{t_0}^{t} \exp(s/T_i)\,dW_i(s), \label{eq:gamma_exa}\\
& \quad \Gamma _i(t) = \frac{1}{\tau_p}\exp(-t/\tau_p)
  \int _{t_0}^{t}\exp(s/\tau_p)\,\gamma _i(s)\,ds, \label{eq:Gamma_exa}\\
& \quad \Omega _i(t) = \int _{t_0}^{t}\Gamma
_i(s)\,ds. \label{eq:Omega_exa} \\ \notag \\
& \text{\underline{By resorting to stochastic integration by parts, $\gamma
_i(t),\;\Gamma _i(t),\;\Omega _i(t)$ can be written:}}\notag \\
& \quad \gamma _i(t) = \Check{B}_i\,\exp(-t/T_i)
\,I_{1,i}, \label{eq:gammaN_exa}\\
& \quad \Gamma _i(t) = \theta_i \,\Check{B}_i\,
 [\exp(-t/T_i)\,I_{1,i} -\exp(-t/\tau_p)\,I_{2,i}], \label{eq:GammaN_exa}\\
& \quad \Omega _i(t) = \theta_i \,\Check{B}_i\,
   \{ (T_i-\tau_p)\,I_{3,i} 
 -[T_i \exp(-t/T_i)\,I_{1,i} -
   \tau_p \exp(-t/\tau_p) \,I_{2,i}]\}, \label{eq:OmegaN_exa} \\ 
& \text{with} \quad I_{1,i} = \int_{t_0}^{t}\exp(s/T_i)\,dW_i(s),
\quad I_{2,i}= \int_{t_0}^{t}\exp(s/\tau_p)\,dW_i(s), \quad I_{3,i}=\int_{t_0}^{t}dW_i(s).\notag
\end{align}
\hrule
\label{tab:exa}
\end{table}
%
%
\begin{table}[htbp]
\caption{Derivation of the covariance matrix for constant coefficients.}
\hrule
\begin{align}
& \lra{\gamma _i^2(t)} = \Check{B}_i^2 \,\frac{T_i}{2}
  \left[1-\exp(-2\Delta t/T_i)\right] \quad \text{where} \quad
  \Check{B}_i^2 = B_{ii}^2 \label{eq:gama2} \\ \notag 
& \lra{\Gamma _i^2(t)} =
    \Check{B}_i^2 \,\theta_i ^2
   \left\{\frac{T_i}{2}[1-\exp(-2\Delta t/T_i)]
  - \frac{2\tau_p T_i}{T_i+\tau_p}
    [1-\exp(-\Delta t/T_i)\exp(-\Delta t/\tau_p)] \right . \notag \\
& \hspace*{11mm} + \left . \frac{\tau_p}{2}[1-\exp(-2\Delta t/\tau_p)] \right\}
  \label{eq:Gama2} \\ \notag 
& \frac{1}{\Check{B}_i^2 \,\theta_i ^2}\lra{\Omega _i^2(t)} =
  (T_i-\tau_p)^2\Delta t + \frac{T_i^3}{2}[1-\exp(-2\Delta t/T_i)]
  + \frac{\tau_p^3}{2}[1-\exp(-2\Delta t/\tau_p)] \notag \\
&\hspace*{19mm} - 2T_i^2(T_i-\tau_p)[1-\exp(-\Delta t/T_i)] 
+ 2\tau_p^2(T_i-\tau_p)[1-\exp(-\Delta t/\tau_p)] \notag \\
&\hspace*{19mm} - 2\frac{T_i^2\tau_p^2}{T_i+\tau_p}
    [1-\exp(-\Delta t/T_i)\exp(-\Delta t/\tau_p)] \\ \notag 
& \lra{\gamma _i(t)\,\Gamma _i(t)} =
    \Check{B}_i^2 \, \theta_i \,T_i
\left\{ \frac{1}{2}[1-\exp(-2\Delta t/T_i)]
  - \frac{\tau_p}{T_i+\tau_p}[1-\exp(-\Delta t/T_i)
    \exp(-\Delta t/\tau_p)]\right\} \label{eq:gaGam} \\ \notag 
& \lra{\gamma _i(t)\,\Omega _i(t)} =
  \Check{B}_i^2 \, \theta_i \,T_i
  \left\{(T_i-\tau_p)[1-\exp(-\Delta t/T_i)]
  - \frac{T_i}{2}[1-\exp(-2\Delta t/T_i)] \right . \notag \\
&  \hspace*{17mm} + \left . \frac{\tau_p^2}{T_i+\tau_p}[1-\exp(-\Delta t/T_i)
    \exp(-\Delta t/\tau_p)]\right\} \\ \notag 
& \frac{1}{\Check{B}_i^2 \, \theta_i ^2}
  \lra{\Gamma _i(t)\,\Omega _i(t)} =
  (T_i-\tau_p)\{T_i[1-\exp(-\Delta t/T_i)]-
               \tau_p[1-\exp(-\Delta t/\tau_p)]\}  \notag \\
&  \hspace*{26mm} - \frac{T_i^2}{2}[1-\exp(-2\Delta t/T_i)] 
  - \frac{\tau_p^2}{2}[1-\exp(-2\Delta t/\tau_p)] \notag \\
& \hspace*{26mm} + T_i \tau_p \,[1-\exp(-\Delta t/T_i)\exp(-\Delta t/\tau_p)] \\ 
& \text{\underline{The stochastic integrals $\gamma _i^n,\;\Omega
_i^n,\;\Gamma_i^n$ are simulated by:}}\notag \\
& \quad \gamma_i^n = P^i_{11}\,{\mc G}_{1,i},\notag \\
& \quad \Omega_i^n = P^i_{21}\,{\mc G}_{1,i}+ P^i_{22}\,{\mc G}_{2,i} \notag \\
& \quad \Gamma_i^n = P^i_{31}\,{\mc G}_{1,i}+ P^i_{32}\,{\mc G}_{2,i}+
                     P^i_{33}\,{\mc G}_{3,i}, \notag\\ 
& \quad \text{where ${\mc G}_{1,i},\;{\mc G}_{2,i},\;{\mc G}_{3,i}$ are
  independent ${\cal N}(0,1)$ random variables.} \notag\\ \notag \\
& \text{\underline{The coefficients
 $P^i_{11},\;P^i_{21},\;P^i_{22},\;P^i_{31},\;P^i_{32},\;P^i_{33}$ are defined
 as:}}\notag\\
& \quad P^i_{11} = \sqrt{\lra{(\gamma_i^n)^2}}, \notag \\
& \quad P^i_{21} = \frac{\lra{\Omega_i^n\gamma_i^n}}{\sqrt{\lra{(\gamma_i^n)^2}}}, 
  \quad P^i_{22} = \sqrt{\lra{(\Omega_i^n)^2}-
  \frac{\lra{\Omega_i^n\gamma_i^n}^2}{\lra{(\gamma_i^n)^2}}},\notag \\
& \quad P^i_{31} = \frac{\lra{\Gamma_i^n\gamma_i^n}}{\sqrt{\lra{(\gamma_i^n)^2}}},
   \quad P^i_{32} = \frac{1}{P^i_{22}}(\lra{\Omega_i^n\Gamma_i^n}-P^i_{21}P^i_{31}),
   \quad P^i_{33} = \sqrt{\lra{(\Gamma_i^n)^2}-(P^i_{31})^2-(P^i_{32})^2)}.\notag
\label{eq:GamOme}
\end{align}
\hrule
\label{tab:matcov_exa}
\end{table}

\clearpage
\bibliography{bibliography}


%
%

%


\end{document}